\documentclass[acmtog,anonymous=false,review=false,screen]{acmart}

\usepackage{booktabs} 

\usepackage{geometrycollective}

\usepackage{dsfont} 

\usepackage[ruled]{algorithm2e} 

\SetAlFnt{\small}
\SetAlCapFnt{\small}
\SetAlCapNameFnt{\small}
\SetAlCapHSkip{0pt}
\IncMargin{-\parindent}

\acmJournal{TOG}

\usepackage[normalem]{ulem}

\setcopyright{rightsretained}

\received{October 2022}

\begin{teaserfigure}
  \centering
  \includegraphics[width=\linewidth]{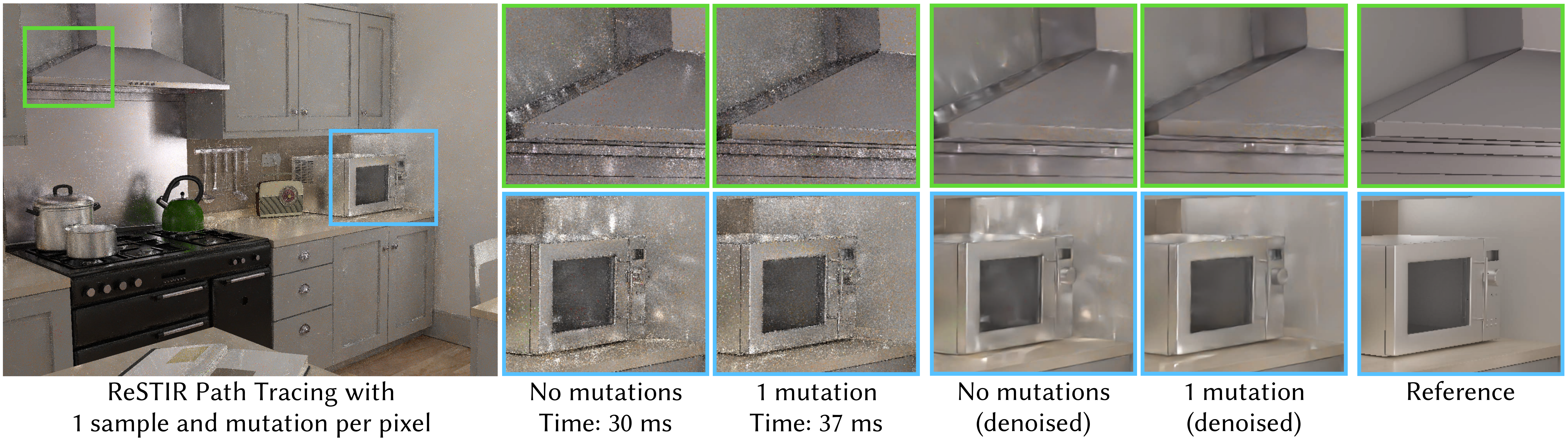}
  \caption{A single sample per pixel (spp) comparison of indirect illumination rendered using ReSTIR Path Tracing (PT) \citep{Lin:2022:GRIS} with and without our sample mutations. By performing even a single mutation per sample, our approach can suppress correlation artifacts that may arise within ReSTIR samplers due to unchecked spatiotemporal reuse. Mutations improve visual fidelity of both rendered and denoised results (with the OptiX denoiser \citep{Nvidia:2017:Optix}) while leaving mean squared error unchanged. \label{fig:Teaser}}
  \label{fig:teaser}
\end{teaserfigure}

\def\equationautorefname~#1\null{%
  Equation~#1\null
}

\usepackage{xr}
\makeatletter
\newcommand*{\addFileDependency}[1]{
  \typeout{(#1)}
  \@addtofilelist{#1}
  \IfFileExists{#1}{}{\typeout{No file #1.}}
}
\makeatother

\newcommand*{\myexternaldocument}[1]{
    \externaldocument{#1}
    \addFileDependency{#1.tex}
    \addFileDependency{#1.aux}
}

\definecolor{killcolor}{rgb}{0.5,0.0,0.0}
\definecolor{proposecolor}{rgb}{0.0,0.8,0.0}

\newcommand{\phat}{\hat{p}}

\newcommand{\diff}{\text{d}}

\renewcommand{\vec}[1]{\ensuremath{\boldsymbol{\mathbf{#1}}}}
\newcommand{\x}{\vec{x}}
\newcommand{\y}{\vec{y}}

\newcommand{\xbar}{\bar{\x}}
\newcommand{\ybar}{\bar{\y}}
\newcommand{\ubar}{\bar{\vec{u}}}
\newcommand{\jdet}[2]{\left|\frac{\partial{#1}}{\partial{#2}}\right|}

\myexternaldocument{supplement}

\begin{document}
\title{Decorrelating ReSTIR Samplers via MCMC Mutations} 

\author{Rohan Sawhney}
\email{rohansawhney@cs.cmu.edu}
\affiliation{%
  \institution{Carnegie Mellon University}
  \country{USA}
}
 
\author{Daqi Lin}
\email{daqil@nvidia.com}
\affiliation{%
  \institution{NVIDIA}
  \country{USA}
}

\author{Markus Kettunen}
\email{mkettunen@nvidia.com}
\affiliation{%
  \institution{NVIDIA}
  \country{Finland}
}

\author{Benedikt Bitterli}
\email{bbitterli@nvidia.com}
\affiliation{%
  \institution{NVIDIA}
  \country{USA}
}

\author{Ravi Ramamoorthi}
\email{ravir@cs.ucsd.edu}
\affiliation{%
  \institution{UC San Diego and NVIDIA}
  \country{USA}
}

\author{Chris Wyman}
\email{cwyman@nvidia.com}
\affiliation{%
  \institution{NVIDIA}
  \country{USA}
}

\author{Matt Pharr}
\email{mpharr@nvidia.com}
\affiliation{%
  \institution{NVIDIA}
  \country{USA}
}

\begin{abstract}
    Monte Carlo rendering algorithms often utilize correlations between pixels to improve efficiency and enhance image quality. For real-time applications in particular, repeated \emph{reservoir resampling} offers a powerful framework to reuse samples both spatially in an image and temporally across multiple frames. While such techniques achieve equal-error up to $100\times$ faster for real-time direct lighting \citep{Bitterli:2020:ReSTIR} and global illumination \citep{Ouyang:2021:ReSTIRGI, Lin:2021:ReSTIRVol}, they are still far from optimal. For instance, unchecked spatiotemporal resampling often introduces noticeable correlation artifacts, while reservoirs holding more than one sample suffer from impoverishment in the form of duplicate samples. We demonstrate how interleaving \emph{Markov Chain Monte Carlo (MCMC)} mutations with reservoir resampling helps alleviate these issues, especially in scenes with glossy materials and difficult-to-sample lighting. Moreover, our approach does not introduce any bias, and in practice we find considerable improvement in image quality with just a single mutation per reservoir sample in each frame.
\end{abstract}

%
%

\ccsdesc[500]{Computing methodologies~Ray Tracing}

%
%

\keywords{real-time rendering, resampled importance sampling, weighted reservoir sampling, Markov chain Monte Carlo}



\maketitle

\section{Introduction}
\label{sec:Intro}

The efficiency of rendering algorithms often hinges on their ability to effectively evaluate similar integrals by reusing samples across pixels \citep{Ward:1988:Radiance, Lafortune:1993:Bidirectional, Jensen:1996:Photon, Veach:1997:MLT}. In real-time path tracing, sample reuse becomes more critical since tracing rays is computationally intensive even on high-end consumer GPUs \citep{Kilgariff:2018:Nvidia}. Moreover, while existing denoisers drastically improve image quality even at low sample counts~\citep{Chaitanya:2017:Denoising, Schied:2017:FilteringA, Schied:2018:FilteringB, Pawel:2019:ReLAX, Nvidia:2022:NRD}, they are unable to reconstruct features missing from their input samples. Thus, sample reuse is often the only means to improve sampling quality given limited computational budgets. Compared to methods that generate independent samples, reuse is also at times the only practical approach available to render challenging scenes with caustics and tricky lighting \citep{Hachisuka:2009:SPPM, Veach:1997:MLT}.

Recent sampling algorithms for real-time ray tracing achieve massive speedups in scenes with complex illumination by sharing samples spatially within an image and temporally across frames~\citep{Bitterli:2020:ReSTIR, Ouyang:2021:ReSTIRGI, Lin:2021:ReSTIRVol, Lin:2022:GRIS}. These so-called \emph{ReSTIR}\footnote{acronym for \textbf{Re}servoir-based \textbf{S}patio-\textbf{T}emporal \textbf{I}mportance \textbf{R}esampling} based techniques select $N$ high-contribution samples from a larger streamed candidate pool of size $M$. They do so by reformulating \emph{resampled importance sampling (RIS)} \citep{Talbot:2005:Importance} in terms of \emph{weighted reservoir sampling (WRS)} \citep{Chao:1982:WRS}. While RIS effectively importance samples candidates in proportion to an \emph{arbitrary} target function (\eg, the integrand of the rendering equation), WRS makes resampling efficient by reducing storage costs from $O(M)$ to $O(N)$. Repeated resampling across pixels then helps distribute important samples over several frames for estimation.

\begin{figure}[t]
    \centering
    \includegraphics[width=\columnwidth]{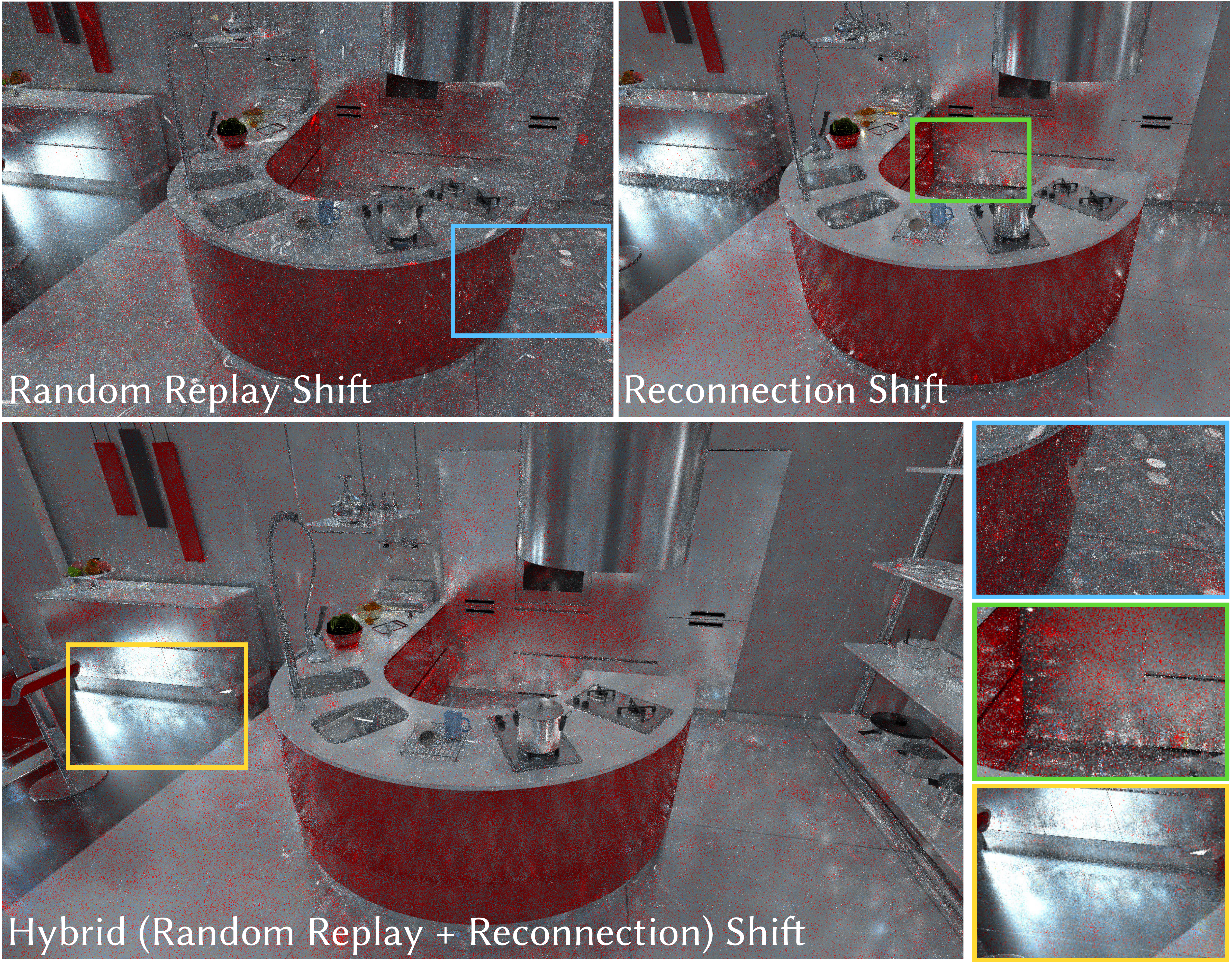}
    \caption{Glossy scenes with difficult-to-sample lighting rendered using ReSTIR PT often contain correlation artifacts irrespective of the selected shift mapping strategy \citep[Section 7]{Lin:2022:GRIS}. Artifacts result from suboptimal importance sampling and over-enthusiastically sharing a few high-contribution samples between pixels.}
    \label{fig:CorrelationArtifacts}
\end{figure}

\begin{figure}[t]
    \centering
    \includegraphics[width=\columnwidth]{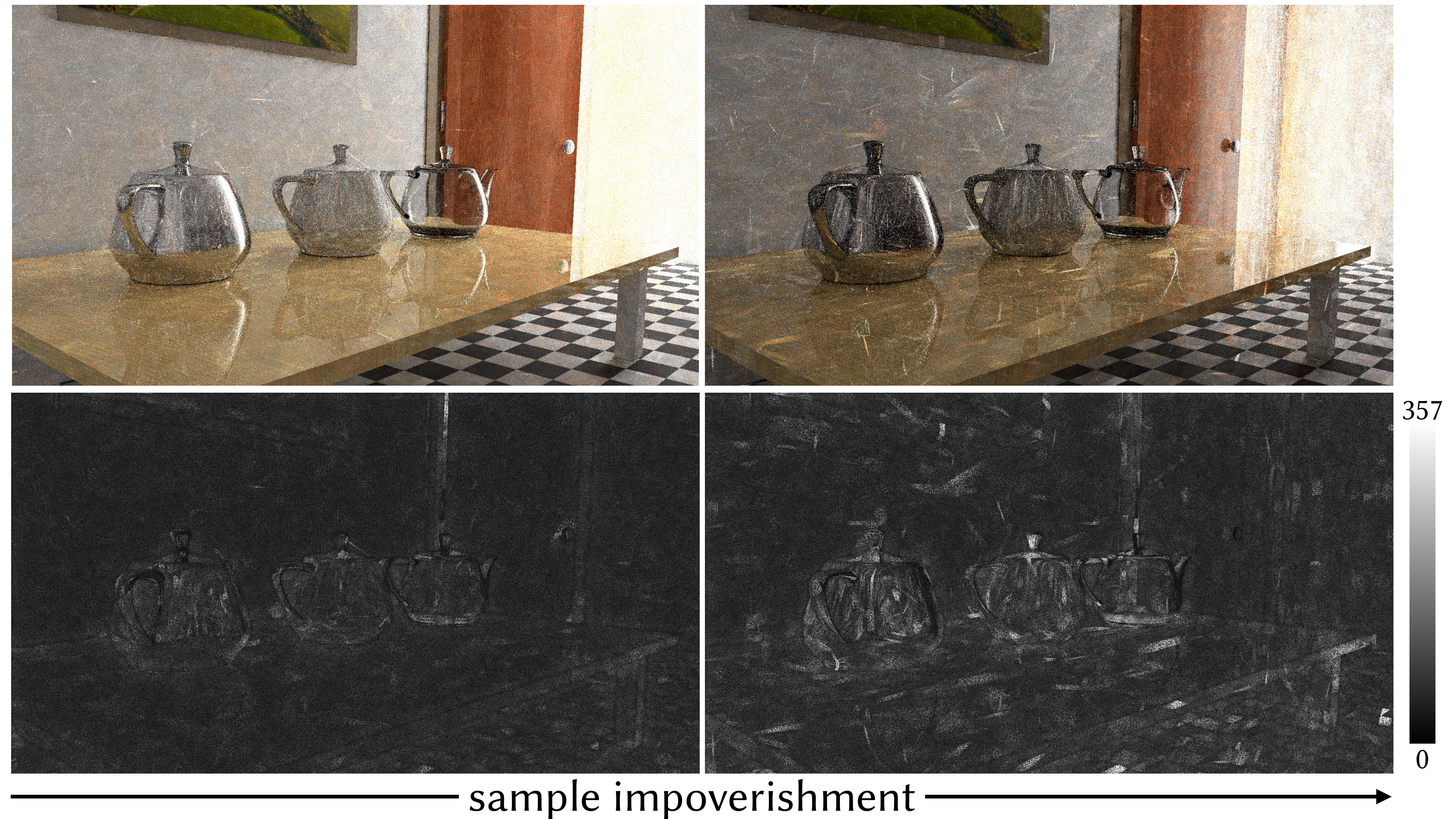}
    \caption{Reservoir resampling suffers from sample impoverishment as it becomes more difficult to sample light-carrying paths. \emph{Top row, left to right:} The Veach Ajar scene rendered using ReSTIR PT (random replay shift) at $1$ spp with the door's angle decreasing. \emph{Bottom row:} Heat maps visualize duplicate samples in $20 \times 20$ pixel neighborhoods. Black represents no duplicates, while white indicates the number of identical samples in a neighborhood.}
    \label{fig:SampleImpoverishment1}
\end{figure}

Though ReSTIR derives impressive efficiency gains from correlated sampling, the benefits of repeated resampling are not indefinite. When only a few high-contribution samples have been identified, iterative spatial reuse creates blotchy artifacts as several pixels reuse the same sample (Figures~\ref{fig:CorrelationArtifacts} and~\ref{fig:SampleImpoverishment1}). Such undersampling artifacts eventually fade away with temporal reuse over several frames, using a parameter to balance between minimizing pixel error and correlations via greater sample reuse (\figref{MCapArtifacts}). However, emphasizing error reduction adds lag under camera movement with dynamically changing lighting and geometry (\secref{SpatiotemporalReuse}), and introduces distracting low-frequency artifacts akin to those in photon mapping \citep{Hachisuka:2009:SPPM}, Metropolis Light Transport (MLT) \citep{Veach:1997:MLT} and Virtual Point Light (VPL) methods \citep{Dachsbacher:2014:ManyLight}.

As spatiotemporal correlations are difficult to quantify, resolving artifacts is challenging. For instance, popular denoisers that compute first- and second-order moments (\eg, \citet{Schied:2017:FilteringA}) are less effective given imprecise variance estimates with correlated samples. For ReSTIR, trying to reduce such artifacts by increasing the reservoir size $N$ is also ineffective, as resampling \emph{with replacement} \citep{Chao:1982:WRS} produces duplicate samples in the presence of strong correlations (see \citet[Figure 19]{Wyman:2021:Rearchitecting}).

Inspired by work on \emph{Sequential Monte Carlo (SMC)} \citep{Doucet:2001:SMC} and \emph{Population Monte Carlo (PMC)} \citep{Cappe:2004:PMC}, we demonstrate that interleaving MCMC mutations with reservoir resampling (\secref{Method}) helps alleviate correlations and impoverishment, especially in scenes with glossy materials and difficult lighting. Unlike MLT where mutations drive information sharing across pixels, our mutations instead help mitigate artifacts caused by spatiotemporal reuse. Similar to blue-noise sampling \citep{Mitchell:1987:antialiased,Georgiev:2016:BlueNoise, Heitz:2019:BlueNoise}, these mutations produce images with better visual fidelity without necessarily reducing error (Figures~\ref{fig:ImpactOnVariance} and ~\ref{fig:DifficultLightPathsMutations}). Our approach highlights the complementary strengths of resampling and mutations for real-time rendering: resampling reuses samples with large contributions proportional to a pixel's target distribution, while mutations diversify the resampled population by locally perturbing samples in proportion to the same target distribution. Furthermore, like \citet{Veach:1997:MLT}'s bias elimination strategy for MLT, we show that resampling eliminates the need for \emph{any} burn-in period with \emph{Metropolis--Hastings (MH)} mutations \citep{Metropolis:1953:MH, Hastings:1970:MH} (\secref{MetropolisHastings}, \appref{StartupBiasElimination}). This drives considerable image quality improvements from even a single mutation per frame for each reservoir sample (Figures~\ref{fig:EqualTimeComparison}, ~\ref{fig:ImpactOnVariance} and ~\ref{fig:MCapAblation}).

From an implementation perspective, our approach requires only simple additions to existing ReSTIR algorithms (see \algref{MutateSample})---
we mutate reservoir samples using Metropolis--Hastings and an appropriate target function every frame after temporal reuse. This is immediately followed by an adjustment to each mutated sample's \emph{contribution weight} to maintain \emph{detailed balance} and ensure unbiased estimation. Overall, our contributions include:
\begin{itemize}
    \item Demonstrating how to incorporate MCMC mutations within ReSTIR to address the pitfalls of unchecked spatiotemporal sample reuse with resampling.
    \item Showing how to correctly adjust the RIS weight of mutated samples in an unbiased fashion for further resampling.
    \item Situating ReSTIR in the broader paradigm of techniques that jointly apply resampling and mutations to sampling problems, such as MLT, SMC and PMC (see \tabref{RelatedWorkTable}).
\end{itemize}

We start with the key building blocks of our approach in the next section, and postpone discussion about related work to \secref{RelatedWork} for better context when comparing with our method.


\section{Background}
\label{sec:Background}

The rendering equation \citep{Kajiya:1986:Rendering} gives the outgoing radiance $L_{\text{out}}$ leaving a point $y$ in the direction $\omega$. Expressed as an integral over directions, it is
\begin{equation}
    \label{eq:RenderingEquation}
    L_{\text{out}}(y, \omega) = L_{\text{e}}(y, \omega) + \int_{S^2} L_{\text{in}}(y, \omega_i)\ \rho(y, \omega, \omega_i)\ |\text{cos}\ \theta_i|\ \diff\omega_i.
\end{equation}
Here $L_{\text{e}}$ is the emitted radiance, $L_{\text{in}}(y, \omega_i)$ is the incoming radiance from the direction $\omega_i$, $\rho(y, \omega, \omega_i)$ is the BSDF and $\theta_i$ is the angle between $\omega_i$ and the surface normal at $y$. Absent participating media, the incident radiance $L_{\text{in}}$ is defined recursively as $L_{\text{in}}(y, \omega_i) = L_{\text{out}}(t(y, \omega_i), -\omega_i)$; the function $t(y, \omega_i)$ returns the point on the closest surface from $y$ in direction $\omega_i$. 
Integrating over the sphere of directions $S^2$ then gives the total radiance scattered towards $\omega$; this integral can be estimated with Monte Carlo as
\begin{equation}
    \label{eq:RenderingEquationMC}
    \frac{1}{N}\sum_{i=1}^N \frac{L_{\text{in}}(y, \omega_i)\ \rho(y, \omega, \omega_i)\ |\text{cos}\ \theta_i|}{p(\omega_i)},
\end{equation}
where $p(\omega_i)$ is the probability density function (PDF) with respect to solid angle used to sample the incident directions $\omega_i$.  

As in Kajiya's formulation, sometimes it is more convenient to reformulate \eqref{RenderingEquation} over surfaces. To keep the discussion independent of the choice of formulation, we use $\int_{\Omega} f(x)\ \diff x$ to generically represent the integral we want to evaluate with $\Omega$ as its domain. This integral can likewise be estimated using
\begin{equation}
    \label{eq:MC}
    \widehat{I}_{\text{MC}} := \frac{1}{N} \sum_{i=1}^N \frac{f(x_i)}{p(x_i)},
\end{equation}
where $x_i$ are independent random samples drawn from any \emph{source} PDF $p$ that is non-zero on the support of $f$. In rendering, one often draws samples proportional to individual terms of the rendering equation to reduce variance (\eg, the BSDF $\rho$). To perform even better importance sampling, ReSTIR instead uses RIS to draw samples \emph{approximately} proportional to the product of multiple terms in the integrand (\eg, $L_{\text{in}} \cdot \rho \cdot |\text{cos}\ \theta|$). 

We review RIS and generalized RIS next (Sections~\ref{sec:RIS} and~\ref{sec:GRIS}); \secref{WRS} discusses a streaming RIS implementation via reservoir sampling. \secref{SpatiotemporalReuse} then describes how correlations arise within ReSTIR due to resampling. \secref{MetropolisHastings} discusses the Metropolis--Hastings algorithm we use in \secref{Method} to resolve correlation artifacts.

\subsection{Resampled Importance Sampling (RIS)}
\label{sec:RIS}

RIS \citep{Talbot:2005:Importance, Lin:2022:GRIS} enables \emph{unbiased} estimation and sample generation from a non-negative target function $\phat(x)$ with an unknown normalization factor $\int_{\Omega} \phat(y)\ \diff y$. It does so by rewriting the standard Monte Carlo estimator from \eqref{MC} as
\begin{equation}
    \label{eq:MCWithPhat}
    \frac{1}{N} \sum_{i=1}^N \frac{f(x_i)}{\phat(x_i)}\ \left(\int_{\Omega} \phat(y)\ \diff y\right).
\end{equation}
The normalization factor is estimated by generating $M \geq 1$ candidate samples $\textbf{y} = \{y_1, \ldots, y_M\}$ from a source PDF $q$ that may be suboptimal but easy to sample from (\eg, $q \propto \rho$), yielding
\begin{equation}
    \label{eq:RIS1}
    \frac{1}{N} \sum_{i=1}^N \frac{f(x_i)}{\phat(x_i)}\ \left(\frac{1}{M} \sum_{j=1}^M \frac{\phat(y_j)}{q(y_j)}\right).
\end{equation}
The samples $\textbf{x} = \{x_1, \ldots, x_N\}$ in turn are selected by randomly choosing an index $j \in \{1, \ldots, M\}$, $N$ times, from the candidate pool $\textbf{y}$ with discrete probabilities:
\begin{equation}
    \label{eq:RISProb}
    \mathbb{P}(j\ |\ \textbf{y}) = \frac{w(y_j)}{\sum_{k = 1}^M w(y_k)}.
\end{equation}
Here the \emph{resampling weight} $w$ for each candidate $y_j$ is given by
\begin{equation}
    \label{eq:ResamplingWeight}
    w(y_j) = \frac{1}{M} \phat(y_j) W(y_j),
\end{equation}
where $W := 1/q(y_j)$ is called the \emph{(unbiased) contribution weight} for $y_j$. The selected samples $x_i$ are likewise given contribution weights
\begin{equation}
    \label{eq:ContributionWeight}
    W(x_i) := \frac{1}{\phat(x_i)} \left(\sum_{j=1}^M w(y_j)\right)
\end{equation}
that assume the role of a reciprocal PDF, though they are only unbiased estimates for elements of the resampled set $\textbf{x}$. This is because the parenthesized term for the normalization factor of $\phat$ is itself an estimator that has variance. Each $x_i\!\in\!\textbf{x}$ is also distributed only approximately in proportion to $\phat$ (\ie, $\phat$ is sampled perfectly only in the limit as $M \rightarrow \infty$). Since we resample \emph{with replacement}, the set $\textbf{x}$ can contain duplicate samples, which reflects that samples are selected in proportion to $\phat$. With this setup,
\citet{Talbot:2005:RIS}
shows that the RIS estimator
\begin{equation}
    \label{eq:RIS2}
    \widehat{I}_{\text{RIS}} := \frac{1}{N} \sum_{i=1}^N f(x_i) W(x_i)
\end{equation}
is unbiased as long as $\phat$ and $q$ are non-zero on the support of $f$, \ie,
\begin{equation}
    \label{eq:RIS2Unbiased}
    \mathbb{E}[\widehat{I}_{\text{RIS}}] = \int_{\Omega} f(x)\ \diff x.
\end{equation}

\paragraph{Combining with Multiple Importance Sampling (MIS)} There are often several reasonable sampling strategies available in rendering, \eg, BSDF or light sampling. MIS~\citep{Veach:1995:MIS} allows multiple strategies to be combined robustly within RIS~\citep{Talbot:2005:RIS}. When each candidate $y_j$ has its own source PDF $q_j$, then MIS weights generalize the parenthesized term in \eqref{RIS1} with
\begin{equation}
    \label{eq:MIS}
    \sum_{j=1}^M m_j(y_j) \frac{\phat(y_j)}{q_j(y_j)}.
\end{equation}
Here, $m_j \geq 0$ is the MIS weight for the $j$th sampling technique. These weights must form a partition of unity, \ie, $\sum_{j=1}^M m_j(y) = 1$. A common choice is the \emph{balance heuristic} $m_j(y) = q_j(y)/\sum_{k=1}^M q_k(y)$ \citep{Veach:1995:MIS}. With MIS, the resampling weight in \eqref{ResamplingWeight} becomes:
\begin{equation}
    \label{eq:ResamplingWeightMIS}
    w(y_j) = m_j(y_j) \phat(y_j) W(y_j), \quad\text{where}\quad W(y_j) = \frac{1}{q_j(y_j)}.
\end{equation}
Notice we recover $m_j = 1/M$ when source PDFs are the same for each sample $y_j$. MIS weights play an important role in ReSTIR. Beyond reducing noise in the resampling weights, they also remove bias when the supports of the source and target distributions do not match integrand $f$'s support (see Section 4 in \citet{Bitterli:2020:ReSTIR} and Section 5 in \citet{Lin:2022:GRIS} for further details). 

In practice, using RIS with the balance heuristic is costly, as all sampling strategies (\ie, the source PDFs) must be evaluated for each candidate sample $y_j$. \citet[Chapter 9.1.3]{Bitterli:2022:Correlations} provides a similarly robust but more efficient heuristic called \emph{Pairwise MIS}, which only requires $O(M)$ PDF evaluations over the entire candidate pool. We use pairwise MIS when the number of sampling strategies $M$ is greater than $2$ (\eg, during spatial resampling in ReSTIR; see \secref{SpatiotemporalReuse}); otherwise we use the balance heuristic.

\subsection{Generalized Resampled Importance Sampling (GRIS)}
\label{sec:GRIS}

So far we assumed the resampling inputs $y_j \sim q_j$ share a common integration domain $\Omega$ with integrand $f$. This assumption may no longer hold when reusing spatially or temporally across an image (as in ReSTIR), and depends on the integral formulation used for the rendering equation. For instance, ReSTIR applied to global illumination \citep{Ouyang:2021:ReSTIRGI, Lin:2022:GRIS} generates samples from PDFs with respect to solid angle. 
Reuse across pixels therefore requires a change of integration domain, necessitating a correction term in the resampling weights \citep[Equation 11]{Ouyang:2021:ReSTIRGI}.  ReSTIR for direct lighting \citep{Bitterli:2020:ReSTIR} instead integrates over the surface of all lights, ensuring $\Omega$ is fixed across samples.

Recent work by \citet{Lin:2022:GRIS} generalizes RIS to use candidate samples $y_j$ originating from different domains $\Omega_j$. It achieves this via \emph{shift mapping}, \ie, a bijective transformation of samples from one pixel to corresponding samples on another pixel \citep{Lehtinen:2013:GDMLT}. In particular, if $\Omega$ denotes the domain of integration for $f$, and $S_j: \Omega_j \rightarrow \Omega$ are shifts that map $y_j \in \Omega_j$ to the modified sample $y'_j \in \Omega$, then the resampling weight for $y_j$ becomes
\begin{equation}
    \label{eq:ResamplingWeightShift}
    w(y_j) = m_j(y'_j) \phat(y'_j) W(y_j) \cdot \left|\frac{\partial y'_j}{\partial y_j}\right|,
\end{equation}
where the Jacobian determinant $|\partial y'_j / \partial y_j|$ accounts for the change of integration domain from $\Omega_j$ to $\Omega$. (Jacobians also appear in MIS weights $m_j$; see \appref{MISTemporal}). 
The rest of the RIS procedure in \secref{RIS} remains unchanged---substituting these resampling weights to \eqref{ContributionWeight} provides the contribution weight for the selected $y'_j$. 

Various shift mappings have been proposed to maximize the similarity between $y'_j$ and $y_j$ such that $|\partial y'_j / \partial y_j| \approx 1$ \citep[Section 3]{Hua:2019:GDPTSurvey}. We describe the shift mappings we use in \secref{Implementation}.

\subsection{Weighted Reservoir Sampling (WRS)}
\label{sec:WRS}

WRS \citep{Chao:1982:WRS} facilitates efficient RIS implementations using a single pass over elements in a stream $\{y_1, \ldots, y_M\}$ to select a random sample. As in \secref{RIS}, each stream element has an associated resampling weight $w$. The basic idea is to process the stream one element at a time, and to select---from the $m < M$ elements processed so far---a sample $y_j$ with probability $w(y_j)/\sum_{k=1}^m w(y_k)$. The next stream element $y_{m+1}$ then replaces $y_j$ with probability $w(y_{m+1})/\sum_{k=1}^{m+1} w(y_k)$. The stream length $M$ need not be known ahead of time, and  WRS can be used to select $N > 1$ samples if needed \citep[Chapter 22.6]{Wyman:2021:WRS}.

WRS reduces the storage needed for resampling to $O(N)$. A lightweight data structure called a \emph{reservoir} is typically used to process the stream and store the selected samples, the stream length $M$ and the weight sum $\sum_{j=1}^M w(y_j)$; see \algref{WRS}.

\begin{algorithm}[t]
\caption{Weighted reservoir sampling ($N = 1$)}
\label{alg:WRS}
\begin{algorithmic}[1]
\algblockdefx[Name]{Class}{EndClass}
    [1][Unknown]{\textbf{class} #1}
    {}
\algtext*{EndClass}
\algblockdefx[Name]{IF}{ENDIF}
    [1][Unknown]{\textbf{if} #1 \textbf{then}}
    {}
\algtext*{ENDIF}

\Class[Reservoir]
    \State $x \gets \varnothing$\Comment{output sample}
    \State $w_{\text{sum}} \gets 0$\Comment{sum of resampling weights}
    \State $M \gets 0$\Comment{number of samples seen so far}
    \State $W \gets 0$\Comment{contribution weight (set in \algref{TemporalReuse})}
    \Function{\textnormal{update}}{$y,\ w$}
        \State $w_{\text{sum}} \gets w_{\text{sum}} + w$
        \State $M \gets M + 1$
        \IF[$\text{rand()} < (w/w_{\text{sum}})$]
            \State $x \gets y$
        \ENDIF
    \EndFunction\label{ReservoirUpdate}
\EndClass
\end{algorithmic}
\end{algorithm}

\subsection{Reservoir-based Spatiotemporal Resampling}
\label{sec:SpatiotemporalReuse}

ReSTIR applies RIS and WRS in a \emph{chained} fashion within and across pixels of an image. The first key idea is to approximately importance sample multiple terms in the rendering equation's integrand through a per-pixel target function $\phat$. The second is to reuse samples from neighboring pixels to exploit the similarity between their target functions. The algorithm performs four steps every frame:

\begin{enumerate}
    \item \emph{(Initial resampling)} Select $N$ samples from a candidate pool of $M$ samples at each pixel. Equations \ref{eq:ResamplingWeightMIS} and \ref{eq:ContributionWeight} provide the resampling and contribution weights for the candidate and selected samples respectively. A reservoir stores the selected samples and their contribution weights.
    \item \emph{(Temporal resampling)} Use \algref{TemporalReuse} to reuse samples across two corresponding pixels in consecutive frames $t$ and $t\!-\!1$. The resampling weight for each sample is computed using the contribution weight already stored in its reservoir.
    \item \emph{(Spatial resampling)} For each pixel, select $k$ random reservoirs from a small spatial neighborhood and merge them into the pixel's reservoir. The is similar to \algref{TemporalReuse} and can be repeated multiple times; for reference see  \citet[Algorithm 4]{Bitterli:2020:ReSTIR} and \citet[Algorithm 2]{Ouyang:2021:ReSTIRGI}.
    \item \emph{(Final shading)} Use \eqref{RIS2} to compute each pixel's color. 
\end{enumerate}

Spatiotemporal reuse gives each pixel access to a large population of samples from its local neighborhood. As a result, ReSTIR quickly finds samples that make large contributions to pixels. Nonetheless, gains from sharing samples are not indefinite, and correlation artifacts may arise from undersampling, imperfect shift mappings, and wrongly set parameters. For instance, performing multiple rounds of spatial resampling with too small a pixel radius can lead to blotchy VPL-like artifacts. This happens when RIS cannot effectively importance sample the integrand, \eg, due to difficult-to-sample lighting. Likewise, inadequately designed shift mappings may introduce geometric singularities into a sample's resampling weight via the Jacobian determinant, causing the sample to be widely reused.

\begin{algorithm}[t]
\caption{Combining two reservoirs for temporal reuse ($N = 1$)}
\label{alg:TemporalReuse}
\begin{algorithmic}[1]
\algblockdefx[Name]{RETURN}{ENDRETURN}
    [1][Unknown]{\textbf{return} #1}
    {}
\algtext*{ENDRETURN}
\algblockdefx[Name]{COMMENT}{ENDCOMMENT}
    [1][Unknown]{\textcolor{commentgreen}{\(\triangleright\)\textit{#1}}}
    {}
\algtext*{ENDCOMMENT}

\Require Reservoirs for pixels $i$ and $j$ from frames $t$ and $t-1$ (resp.), and a cap for the latter's sample count
\Ensure A combined reservoir for frame $t$
\Function{combineTemporalReservoirs}{$i,\ j,\ r_i,\ r_j, M_{\text{cap}}$}
    \COMMENT[Cap confidence weight for $r_j$]\ENDCOMMENT
    \State $r_j.M \gets \min(r_j.M,\ M_{\text{cap}})$

    \COMMENT[Compute resampling weight for sample in $r_i$]\ENDCOMMENT
    \State $x_i \gets r_i.x$
    \State $m_i \gets \text{computeMIS}(x_i,\ \phat_i,\ r_i.M,\ \phat_j,\ r_j.M)$\Comment{\eqref{MISTemporal1}}
    \State $w_i \gets m_i \cdot \phat_i(x_i) \cdot r_i.W$\Comment{\eqref{ResamplingWeightMIS}}
    
    \COMMENT[Shift sample in $r_j$ to pixel $i$ and compute its resampling weight]\ENDCOMMENT
    \State $x_j', |\partial x_j'/\partial x_j| \gets \text{shiftMap}(r_j.x,\ j,\ i)$\Comment{\secref{Implementation}}
    \State $m_j \gets \text{computeMIS}(x_j',\ \phat_j,\ r_j.M,\ \phat_i,\ r_i.M)$\Comment{\eqref{MISTemporal2}}
    \State $w_j \gets m_j \cdot \phat_i(x_j') \cdot r_j.W \cdot |\partial x_j'/\partial x_j|$\Comment{\eqref{ResamplingWeightShift}}
    
    \COMMENT[Combine weighted samples into a single reservoir]\ENDCOMMENT
    \State $\text{Reservoir}\ s$
    \State $s.\text{update}(x_i,\ w_i)$
    \State $s.\text{update}(x_j',\ w_j)$
    \State $s.M \gets r_i.M + r_j.M$
    \State $s.W \gets \frac{1}{\phat_i(s.x)}s.w_{\text{sum}}$\Comment{\eqref{ContributionWeight}}
    \RETURN[$s$]
    \ENDRETURN
\EndFunction
\end{algorithmic}
\end{algorithm}

\begin{figure}[t]
    \centering
    \includegraphics[width=\columnwidth]{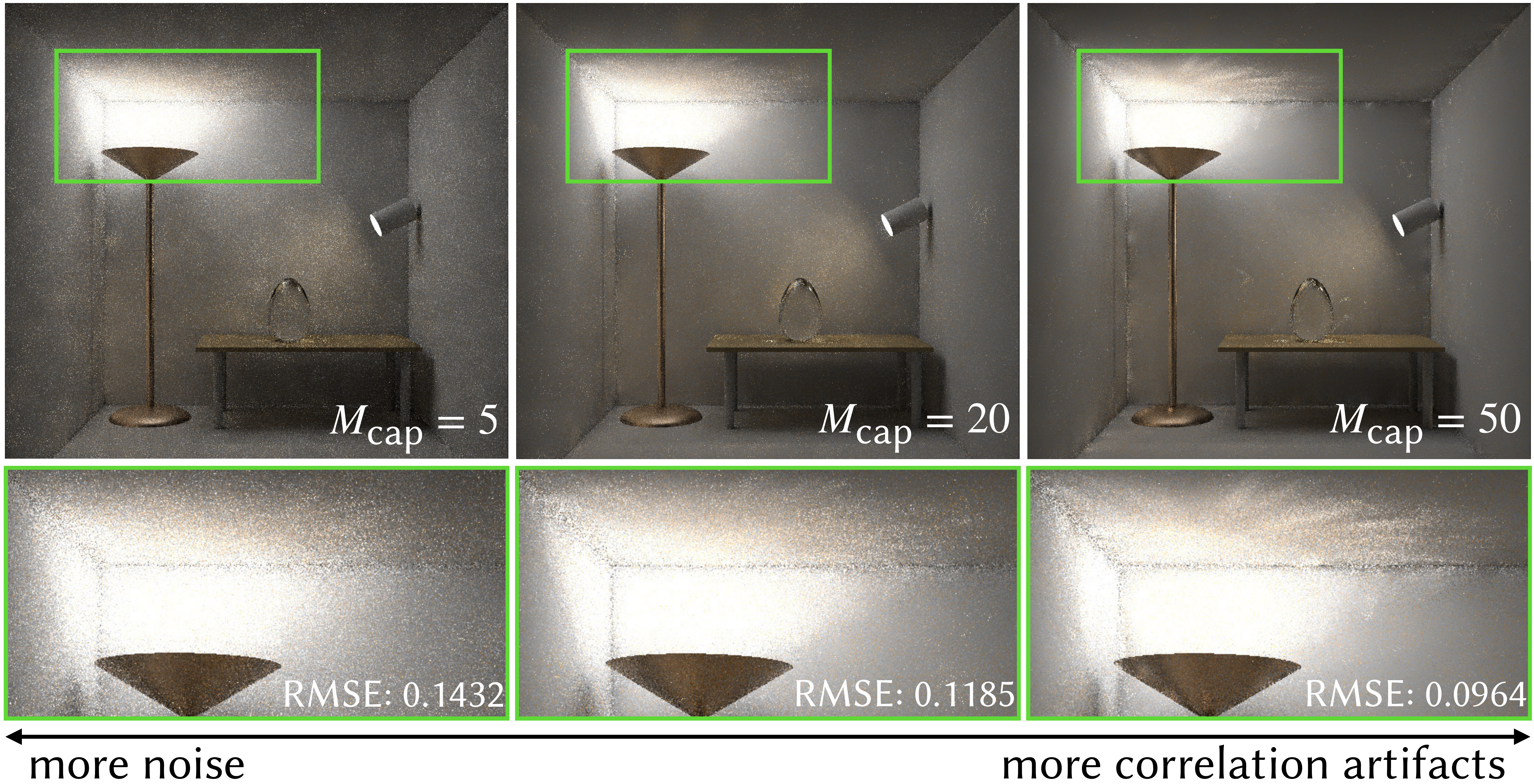}
    \caption{Parameters for ReSTIR sample reuse can be difficult to set in a scene agnostic way. For instance, a large $M_{\text{cap}}$ value introduces correlations (\emph{right}), while a small $M_{\text{cap}}$ inadequately exploits prior samples, leading to noise (\emph{left}). Our approach offers greater leeway in setting parameter values that trade noise for correlation (see \figref{MCapAblation}).}
    \label{fig:MCapArtifacts}
\end{figure}

During temporal resampling, one must cap the stream length or \emph{confidence weight} $M$ of a temporally reused sample (\algref{TemporalReuse}, \emph{line 3}) to guarantee convergence \citep[Section 6.4]{Lin:2022:GRIS}---not doing so results in convergence to a wrong result. Unfortunately, the ideal $M_{\text{cap}}$ cannot always be determined in a scene-agnostic way---small caps inadequately utilize the temporal history and result in higher variance (\citet[Figure 9]{Lin:2022:GRIS}), while large caps increase correlation. In particular, increasing $M_{\text{cap}}$ decreases the relative weight and selection probability of newly proposed samples. As a result, an outlier sample's contribution has to decay for it to match a pixel’s average value. Unfortunately, the outlier may be spread between neighboring pixels before it is eventually replaced. 
This can lead to visible correlation artifacts and sample impoverishment over multiple frames (\figref{MCapArtifacts}). We use the Metropolis-Hastings algorithm, described next, to address these issues in ReSTIR.

\subsection{Metropolis--Hastings (MH)}
\label{sec:MetropolisHastings}

Like RIS, the MH \citep{Metropolis:1953:MH, Hastings:1970:MH} algorithm generates a set of samples distributed proportionally to a non-negative and possibly unnormalized target function $\phat$. While RIS uses resampling to achieve this goal, MH instead constructs a \emph{Markov chain} that has a stationary distribution proportional to $\phat$'s probability density function $\phat/\int_{\Omega} \phat$. In more detail, given an initial sample $x^0 \in \Omega$, MH incrementally constructs a sequence of random samples $x^0, x^1, x^2, ...$ as follows:

\begin{enumerate}
    \item Generate a candidate sample $z^k$ by applying a random \emph{mutation} to the current sample $x^k$ in the chain, \ie, sample $z^k$ from a \emph{proposal density} $T(x^k \rightarrow z^k)$.
    \item Compute an acceptance probability for the candidate $z^k$:
    \begin{equation}
        \label{eq:MHAcceptance}
        a(x^k \rightarrow z^k) := \min\left(1, \frac{\phat(z^k)\ T(z^k \rightarrow x^k)}{\phat(x^k)\ T(x^k \rightarrow z^k)}\right).
    \end{equation}
    \item Set $x^{k+1} = z^k$ with probability $a$; otherwise set $x^{k+1} = x^k$.
\end{enumerate}

The acceptance probability $a(x^k \rightarrow z^k)$ ensures that samples are distributed proportional to the target function $\phat$. The \emph{detailed balance} condition guarantees the existence of the Markov chain's stationary distribution by requiring the transition density between any two sample values to be equal:
\begin{equation}
    \label{eq:DetailedBalance}
    \phat(x^k) T(x^k \rightarrow z^k) a(x^k \rightarrow z^k) = \phat(z^k) T(z^k \rightarrow x^k) a(z^k \rightarrow x^k).
\end{equation}
To generate the correct distribution from all inputs, Markov chains must be \emph{ergodic}. This can be guaranteed easily with mutations that always propose candidate samples over the entire support of $\phat$, i.e., $T(x^k \rightarrow z^k) > 0$ for all $x^k$ and $z^k$ where $\phat(x^k) > 0$ and $\phat(z^k) > 0$. Even with this constraint, there is still much freedom in choosing mutation strategies---\secref{Implementation} describes the strategies we use.

Unlike RIS, MH does not estimate the value of integrals. It does however produce valid samples from its target function which can be used by a secondary estimator such as RIS for estimation (\secref{Method}).

\paragraph{Eliminating start-up bias} MH assumes the initial sample $x^0$ is generated with probability density proportional to $\phat$; using a sample not from this distribution results in \emph{start-up bias}. A typical solution runs the Markov chain for numerous iterations until the initial state is ``forgotten'', \ie, discarding several early samples generated by MH. Sadly, determining the length of this \emph{burn-in} period is tricky, as it depends on the initial sample value and its actual distribution. \citet[Chapter 11.3.1]{Veach:1998:Robust} instead proposed resampling $x^0$ from $M$ candidate samples $\textbf{y} = \{y_1, \ldots, y_M\}$ generated using an easy-to-sample source PDF (much like \secref{RIS}). Equations \ref{eq:RISProb} and \ref{eq:ResamplingWeight} then provide the discrete probabilities and resampling weights (resp.) needed to select a candidate, \ie, $x^0 = y_j$ for some $j \in \{1, \ldots, M\}$. Finally, contributions of mutated samples initialized from $x^0$ are weighted by \eqref{ContributionWeight} to guarantee unbiased estimation. Our method leverages ReSTIR's built-in resampling to automatically avoid start-up bias when performing mutations.


\section{Method}
\label{sec:Method}

RIS improves sample selection from a target distribution when given a large population of candidate samples. ReSTIR amasses a sizable, per-pixel candidate pool for resampling through spatiotemporal reuse, helping it quickly identify high-contribution samples via RIS. However, at times ReSTIR extensively reuses a few samples over multiple frames due to imperfect importance sampling and suboptimal parameters, as it has no mechanism to diversify its existing sample population.

Inspired by Sequential and Population Monte Carlo techniques (\secref{RelatedWork}), we interleave reservoir resampling with MCMC mutations to mitigate correlations and sample impoverishment caused by
spatiotemporal reuse. Our key observation is that mutating reservoir samples with the same target function as RIS helps to quickly decorrelate the resampled population, especially when it contains outliers. In \algref{MutateSample}, we use Metropolis-Hastings to locally perturb the per-pixel temporal samples selected by \algref{TemporalReuse}; interleaving resampling with mutations diversifies the samples ReSTIR shares between pixels. We discuss key aspects of our work next, starting with how to modify mutated samples' contribution weights to guarantee unbiased results.

\begin{figure}[t]
    \centering
    \includegraphics[width=\columnwidth]{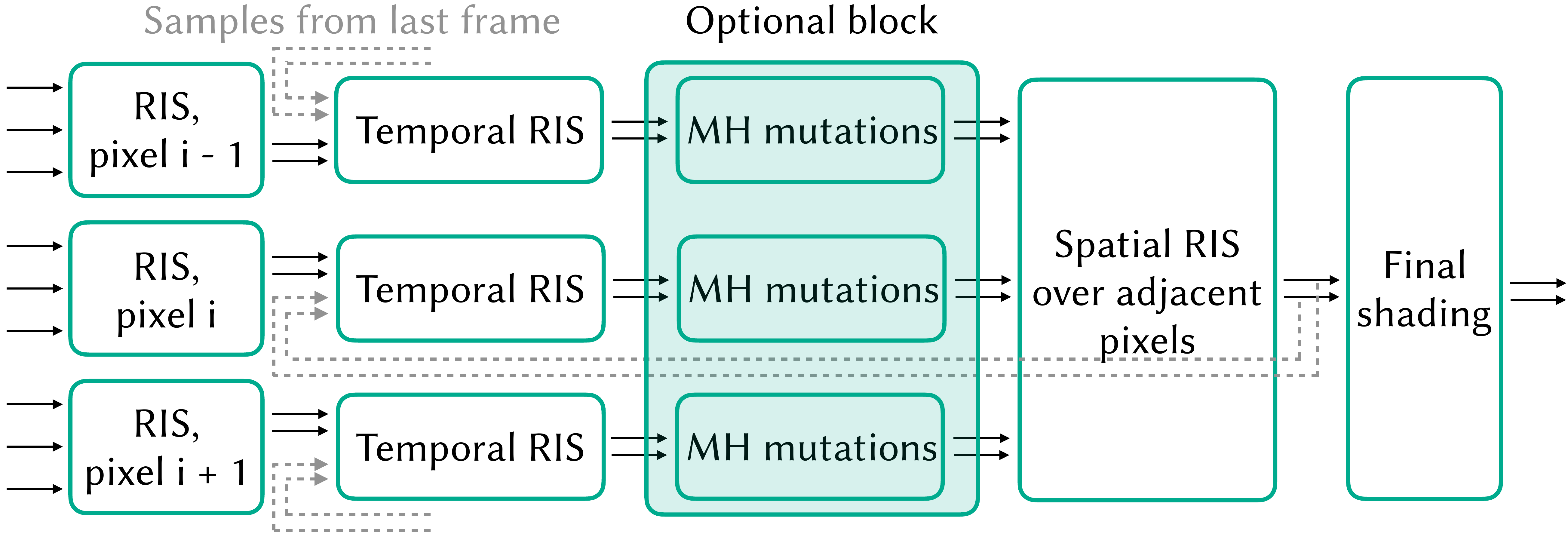}
    \caption{Our approach introduces Metropolis-Hastings mutations as an additional block into the larger ReSTIR algorithm for spatiotemporal sample reuse. Samples are mutated within each pixel after temporal resampling (\algref{TemporalReuse}) to mitigate correlation artifacts and sample impoverishment.}
    \label{fig:ReSTIR}
\end{figure}

\paragraph{Modified contribution weights} A contribution weight $W$ (\eqref{ContributionWeight}) estimates the reciprocal value of the target PDF $\phat/\int_{\Omega} \phat$ that a sample is approximately distributed according to. $W$ is needed to compute resampling weights for combining reservoirs (\algref{TemporalReuse}, \emph{lines 7} and \emph{11}) and to estimate per-pixel shading (\eqref{RIS2}). 

Contribution weights are sample dependent. Thus, a sample that undergoes mutation cannot reuse the weight associated with its original state, \ie, a mutated sample's contribution weight should provide an unbiased estimate for the sample's reciprocal target PDF. Our key contribution is to show that the unbiased contribution weight for any mutated sample $x^k$, from a Markov chain $x^0, ..., x^k, ...$, can be computed via the relation
\begin{equation}
    \label{eq:ModifiedContributionWeight}
    W(x^k) = \frac{\phat(x^0)}{\phat(x^k)} W(x^0).
\end{equation}
\eqref{ModifiedContributionWeight} does not depend on samples between $x^0$ and $x^k$ in the Markov chain and imposes no constraints on computing $W(x^0)$, which can arise from prior resampling, runs of MH, or a mix of the two. This provides flexibility in where and when to mutate samples during ReSTIR (as long as mutations are confined to a given pixel). 

One can get an intuitive feel for \eqref{ModifiedContributionWeight} by substituting in the expression for $W(x^0)$ from \eqref{ContributionWeight}:
\begin{equation}
    \label{eq:ModifiedContributionWeightSubstituion}
    W(x^k)\ =\ \frac{\bcancel{\phat(x^0)}}{\phat(x^k)} \cdot \frac{1}{\bcancel{\phat(x^0)}} \left(\sum_{j=1}^M w(y_j)\right)\ =\ 
    \frac{1}{\phat(x^k)} \left(\sum_{j=1}^M w(y_j)\right)
\end{equation}
Notice that the estimated normalization factor for $\phat$, \ie, the sum of weights $w$, remains unchanged for both the initial and mutated samples $x^0$ and $x^k$. This normalization factor arises via RIS (\eg, \algref{TemporalReuse}, \emph{lines 14-15}) prior to performing mutations. Meanwhile, MH treats the resampling weights as fixed, simply redistributing a reservoir's sample population  proportionally to the per-pixel target function $\phat$. \eqref{ModifiedContributionWeight} then encodes any required correction to a sample's contribution weight to account for the sample mutation.

\begin{algorithm}[t]
\caption{Mutate sample via Metropolis-Hastings}
\label{alg:MutateSample}
\begin{algorithmic}[1]
\algblockdefx[Name]{RETURN}{ENDRETURN}
    [1][Unknown]{\textbf{return} #1}
    {}
\algtext*{ENDRETURN}
\algblockdefx[Name]{COMMENT}{ENDCOMMENT}
    [1][Unknown]{\textcolor{commentgreen}{\(\triangleright\)\textit{#1}}}
    {}
\algtext*{ENDCOMMENT}

\Require Pixel i, reservoir $r_i$ from \algref{TemporalReuse}, and iteration count
\Ensure Reservoir $r_i$ with its sample mutated in proportion to $\hat{p}_i$
\Function{mutateSample}{$i,\ r_i$,\ \text{iters}}
    \State $z \gets \text{metropolisHastings}(r_i.x,\ \phat_i,\ \text{iters})$\Comment{\secref{MetropolisHastings}}
    \State $r_i.W \gets \frac{\phat_i(r_i.x)}{\phat_i(z)}  \cdot r_i.W$\Comment{\eqref{ModifiedContributionWeight}}
    \State $r_i.x \gets z$
    \RETURN[$r_i$]
    \ENDRETURN
\EndFunction
\end{algorithmic}
\end{algorithm}

\paragraph{Start-up bias} \algref{MutateSample} does not require a burn-in period for mutations, even though the samples used to initialize MH are not distributed exactly according to $\phat$. This is because we use the unbiased contribution weights of mutated samples for subsequent steps in ReSTIR, including when computing shading and resampling weights for further reuse. This approach eliminates start-up bias completely for any mutated sample $x^k$ by ensuring
\begin{equation}
    \mathbb{E}[f(x^k) W(x^k)] = \int_{\Omega} f(x)\ \diff x
\end{equation}
for any function $f$. \appref{StartupBiasElimination} provides a formal proof.

Note that avoiding start-up bias does not imply samples generated using MH are well-distributed according to $\phat$. However, since we initialize MH using reservoir samples 
roughly proportional to the target function,
our method does not rely on MH to find important samples (see \figref{DifficultLightPathsMutations})---rather it decorrelates and diversifies outlier samples by mutating them locally in proportion to $\phat$.

\paragraph{When to perform mutations?} Temporal reservoirs often contain stale samples, as ReSTIR assigns higher relative importance to existing samples. We therefore mutate samples output by \algref{TemporalReuse} within each pixel (\figref{ReSTIR}), using the same per-pixel target function as RIS for the current frame.
Mutating samples randomly after temporal resampling diversifies the inputs to spatial resampling, protecting against possibly escalating amounts of sample impoverishment caused by repeated reuse.

Applying \algref{MutateSample} to mutate samples after the initial or spatial resampling steps in ReSTIR (\secref{SpatiotemporalReuse}) is possible but not required.  Like mutations, initial resampling serves to rejuvenate the sample population every frame (by introducing new independent samples into the population). Samples from spatial resampling are stored for future reuse; mutating them proportional to the current target function would cause them to lag by one frame.

Finally, \algref{MutateSample} places no restrictions on MH iteration count. To improve runtime performance, one could adaptively specify mutation counts per pixel (including no mutations) using, for instance, local correlation estimates. We leave development of such heuristics to future work and use a fixed, user-specified number of iterations.

\section{Implementation Details}
\label{sec:Implementation}

We perform mutations for both direct and indirect illumination in ReSTIR using \citet{Kelemen:2002:PSSMLT}'s \emph{primary sample space (PSS)} parameterization. This conveniently allows applying mutations directly to random number sequences used to generate light-carrying paths, while constraining path vertices to remain on the scene manifold. Moreover, it simplifies  use of certain shift mappings in ReSTIR PT, \eg, the \emph{random replay} shift \citep[Section 7.2]{Lin:2022:GRIS}.

In this section, we represent samples with a path vertex notation $\xbar = [\x_0, \x_1, \ldots, \x_k] \in \Omega^k(\mathcal{M})$, with $\Omega^k(\mathcal{M})$ the space of all paths of length $k$ on the scene manifold $\mathcal{M}$ (\eg, $k = 2$ for direct lighting). Each path $\xbar$ is uniquely determined\footnote{
As in \citet{Bitterli:2017:Reversible}, we bijectively
map between paths and their random numbers by padding paths
with extra dimensions.
} by a vector of random numbers $\ubar = [u_0, u_1 \ldots] \in [0, 1]^{O(k)}$. We use $S$ to denote a shift mapping from a base path $\xbar$ in one pixel to an offset path $\ybar$ in another pixel, \ie, $S([\x_0, \x_1, \ldots, \x_k]) = [\y_0, \y_1, \ldots, \y_k]$. Mutated paths and random numbers are represented using primed quantities, \ie, $\ybar'$ and $\ubar'$.

\subsection{Primary sample space}
\label{sec:PSS}
The PSS parameterization reformulates the acceptance probability in \eqref{MHAcceptance} in terms of a \emph{contribution function} $C$ as follows:
\begin{equation}
    \label{eq:MHAcceptancePSS}
    a(\ubar \rightarrow \ubar') := \min\left(1, \frac{C(\ubar')\ T(\ubar' \rightarrow \ubar)}{C(\ubar)\ T(\ubar \rightarrow \ubar')}\right).
\end{equation}
For us $C(\ubar) := \phat(\ybar(\ubar)) / q(\ybar(\ubar))$, where $\phat$ is the per-pixel target function used for temporal resampling and $q$ is the sampling PDF for generating $\ybar$ from the random numbers $\ubar$ \footnote{
The starting unmutated path $\ybar$ for MH could have been generated in ReSTIR from one of many sampling schemes (\eg, light or BSDF sampling), or over multiple rounds of resampling. Here, we do not require the random numbers $\ubar$ that originally generated $\ybar$; Sections~\ref{sec:ReSTIRDIMutations} and ~\ref{sec:ReSTIRPTMutations} discuss the $\ubar$ we use for mutations.} (with mutated path $\ybar'$ likewise generated from $\ubar'$). As suggested by \citet{Kelemen:2002:PSSMLT}, we compute $\ubar'$ by perturbing each element of $\ubar$ with Gaussian noise. We use $s = s_2\exp(-\log(s_2/s_1)U)$ as our perturbation amount with $U \sim [0, 1)$ and $s \in (s_1, s_2]$ .

\subsection{Direct Lighting}
\label{sec:ReSTIRDIMutations}
Our ReSTIR DI mutations perturb the directions of reservoir samples via their random numbers. For direct lighting, path $\ybar = [\y_0, \y_1, \y_2]$ and its PDF $q(\ybar)$ equals $p_{\rho}(\omega) |\text{cos}\ \theta| / |\y_2 - \y_1|^2$, where $p_{\rho}$ is the PDF for importance sampling the BSDF $\rho$, $\omega$ is the unit vector from $\y_1$ to $\y_2$, and $\theta$ is the angle between $\omega$ and the geometric surface normal at $\y_2$. The PDF $q(\ybar')$ is defined analogously for the mutated $\omega'$, pointing from $\y_1$ to $\y'_2$. Random numbers for the starting MH sample $\y_2$ are recovered by inverting the sampling procedure for direction $\omega$ \citep{Bitterli:2017:Reversible}. Since this mutation is symmetric, the transition kernels in \eqref{MHAcceptancePSS} cancel.

\subsection{Indirect Illumination}
\label{sec:ReSTIRPTMutations}
For ReSTIR PT, our mutation strategies build on shift maps. Unlike a mutation, a shift mapping deterministically perturbs a base path $\xbar$ through one pixel into an offset path $\ybar$ through another pixel. For instance, a \emph{random replay shift} reuses the random numbers that generate $\xbar$ to trace $\ybar$. Since tracing a full path is expensive, a \emph{reconnection} is often used to connect the offset path to the base path at a given index $i$, \ie, $\y_{j} = \x_{j}$ for $j \ge i$. Connecting paths immediately with $i = 2$ is called the \emph{reconnection shift}. Compared to random replay, reconnections are often better at producing paths with similar contributions for diffuse surfaces. But reconnecting $\y_{i-2}, \y_{i-1}$ to $\x_i$ on a glossy surface can introduce paths with near-zero throughput, or introduce geometric singularities when $\y_{i-1}$ and $\x_i$ are too close.

\begin{figure}[t]
    \centering
    \includegraphics[width=\columnwidth]{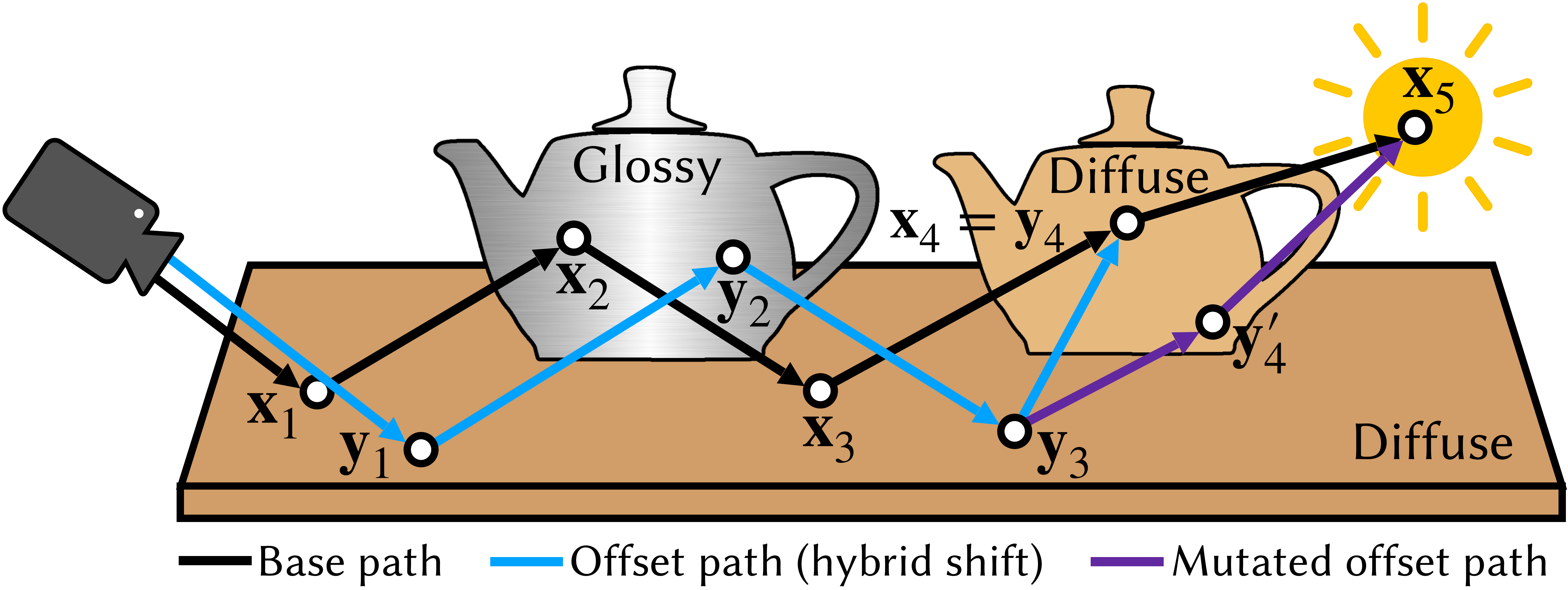}
    \caption{The hybrid shift in ReSTIR PT reconnects the offset path to the base path when it encounters two consecutive diffuse vertices $\x_3, \x_4$; prior to that it reuses random numbers from the base path to trace rays. Our mutation strategy perturbs the reconnection vertex $\y_4$ in the offset path.}
    \label{fig:ShiftMappingMutation}
\end{figure}

We use \citeauthor{Lin:2022:GRIS}'s [\citeyear{Lin:2022:GRIS}] \emph{hybrid} shift strategy (see \figref{ShiftMappingMutation}) to evaluate mutations in ReSTIR PT. This shift mapping postpones reconnection using random replay until certain connectability conditions are met (\eg, surface roughness and distance between vertices).

\paragraph{Mutation strategies} As with direct lighting, one way to mutate a path is to perturb the random numbers used to generate it. Like a random replay shift, this approach expensively requires tracing a full path for each proposed mutation (which may be rejected).

A more computationally efficient approach mutates the offset path with random replay up to the reconnection vertex $\y_i = \x_i$, and then connects to the base path starting at $\x_{i+1}$ instead. We observe this mutation strategy is not only faster, but also has higher acceptance ($70\%$ vs. $40\%$ on the scene from \figref{teaser}) as it minimizes changes to path geometry. Moreover, its paths have similar contributions to the offset paths it mutates. Note that mutating path vertices with random replay until the reconnection to $\x_{i+1}$ can cause connectability conditions for the hybrid shift to fail. We reject such mutated samples by defining their transition PDF to be 0. 

Taking a step further, our final strategy mutates only the reconnection vertex $\y_i$ (\figref{ShiftMappingMutation}) while keeping the rest of the offset path unchanged, i.e., $[\y'_0, \ldots, \y'_k] =$ $[\y_0, \y_1, \ldots, \y_{i-1}, \y'_i, \x_{i+1}, \ldots, \x_k]$, where $\y_{i-1}$ connects to $\y'_i$ with mutated random numbers. We found this strategy only slightly less effective at reducing correlations. It is, however, significantly faster when performing multiple mutations, as only rays from $\y_{i-1}$ to $\y'_i$ and $\y'_i$ to $\x_{i+1}$ need to be traced. We use this mutation strategy to generate results in \secref{ResultsAndDiscussion}, unless otherwise noted. 

Finally, note that the transition kernels $T(\ubar' \rightarrow \ubar)$ and $T(\ubar \rightarrow \ubar')$ are no longer symmetric when offset paths contain a reconnection vertex. In \appref{TransitionKernelReconnection}, we show that their ratio equals:
\begin{equation}
    \label{eq:TransitionKernelRatio}
    \frac{T(\ubar' \rightarrow \ubar)}{T(\ubar \rightarrow \ubar')} = \frac{|\text{cos}\ \theta'|}{|\text{cos}\ \theta|} \frac{|\y_{i+1} - \y_{i}|^2}{|\y_{i+1} - \y'_{i}|^2} \frac{p(\omega_{i-1}', \omega_i')}{p(\omega_{i-1}, \omega_i)} \frac{p(\omega_i', \omega_{i+1})}{p(\omega_i, \omega_{i+1})},
\end{equation}
where $\omega_{i-1}, \omega_i$ and $\omega_{i+1}$ are unit vectors from $\y_{i-1}$ to $\y_i$, $\y_i$ to $\y_{i+1} (= \x_{i+1})$ and $\y_{i+1}$ to $\y_{i+2} (= \x_{i+2})$ respectively, $\theta$ is the angle between $\omega_{i}$ and the surface normal at $\y_{i+1}$, and $p$ is the solid angle PDF used to sample an outgoing direction. Primed quantities are defined similarly. Any mutations applied to random numbers for the subpath $[\y_0, \y_1, \ldots, \y_{i-1}]$ do not factor in the ratio as they are symmetric.

\paragraph{Reservoir storage} \citet[Section 8.2]{Lin:2022:GRIS} note ReSTIR PT stores additional data in the reservoir from \algref{WRS}, specifically a seed for random replay and the resampled path's reconnection vertex. For the first two mutation strategies above, we need the path's entire random number sequence since PSS mutations transform this sequence---as a result, it cannot be regenerated from its original seed. This increases the reservoir size as path length grows. Luckily, our final mutation strategy avoids this overhead, only mutating random numbers that sample $\y'_i$ from fixed offset vertex $\y_{i-1}$. As in ReSTIR DI, we recover random numbers for $\y_i$ by inverting the sampling of direction $\y_i - \y_{i-1}$. The only additional information we store is the offset vertex $\y_{i+1}$ (which connects to mutated vertex $\y'_i$).

\section{Results and Discussion}
\label{sec:ResultsAndDiscussion}

\begin{figure}[t]
    \centering
    \includegraphics[width=\columnwidth]{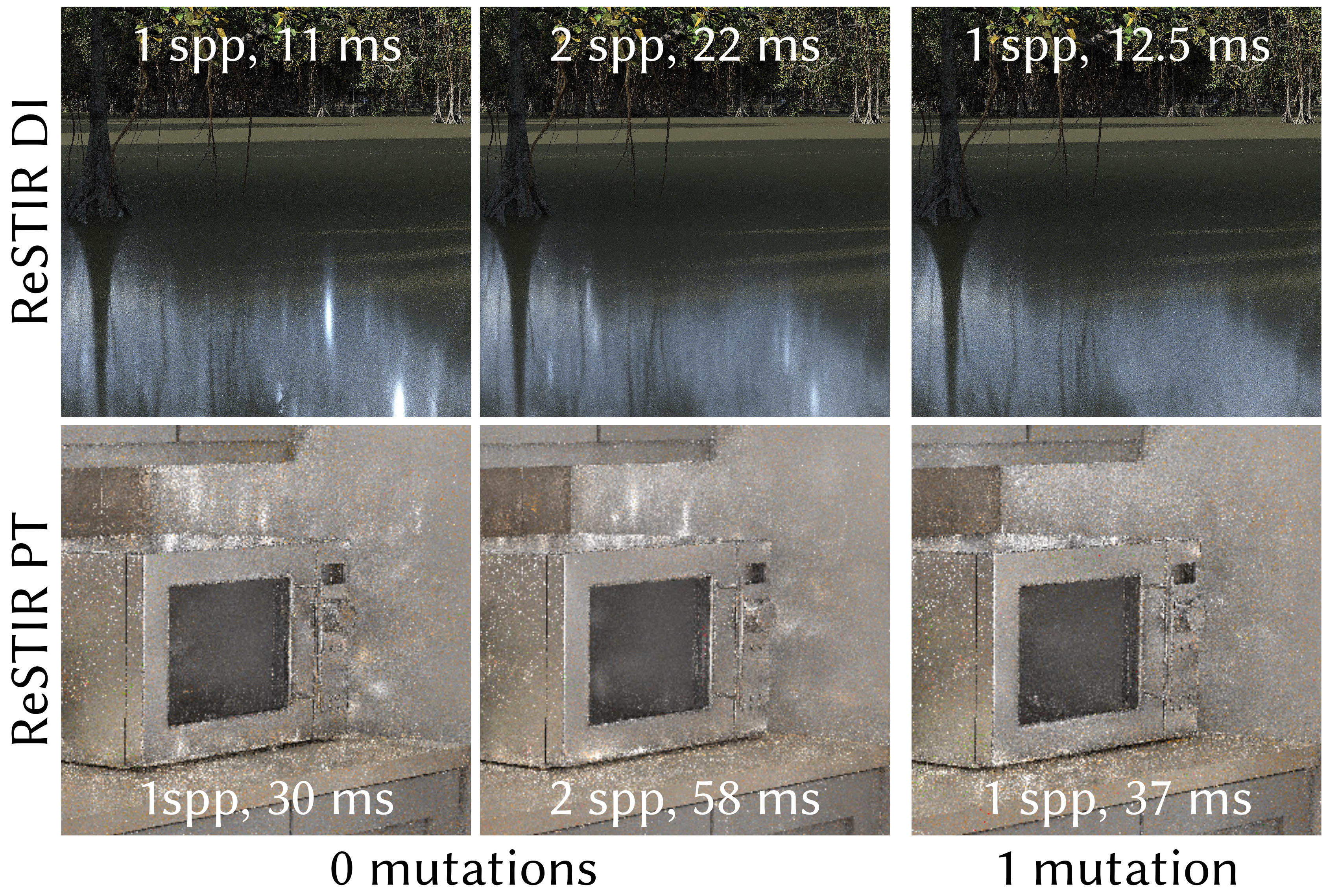}
    \caption{Correlation artifacts often do not disappear simply by using more samples, justifying the overhead of performing mutations. \label{fig:EqualTimeComparison}}
\end{figure}

We prototyped our method in the open-source Falcor rendering framework \citep{Kallweit:2022:Falcor}.
All results use a GeForce RTX 3090 GPU at 1920 $\times$ 1080 resolution. Our implementation uses the settings (\eg, spatial neighborhood size and reuse radius) proposed in \citet{Bitterli:2020:ReSTIR} and \citet{Lin:2022:GRIS} for direct and indirect illumination, with the exception of $M_{\text{cap}} = 50$ in our ReSTIR PT tests. Our supplementary videos show 1 spp results for all our scenes; \tabref{CovarianceNumbers} gives single frame timings.

\begin{figure}[t]
    \centering
    \includegraphics[width=\columnwidth]{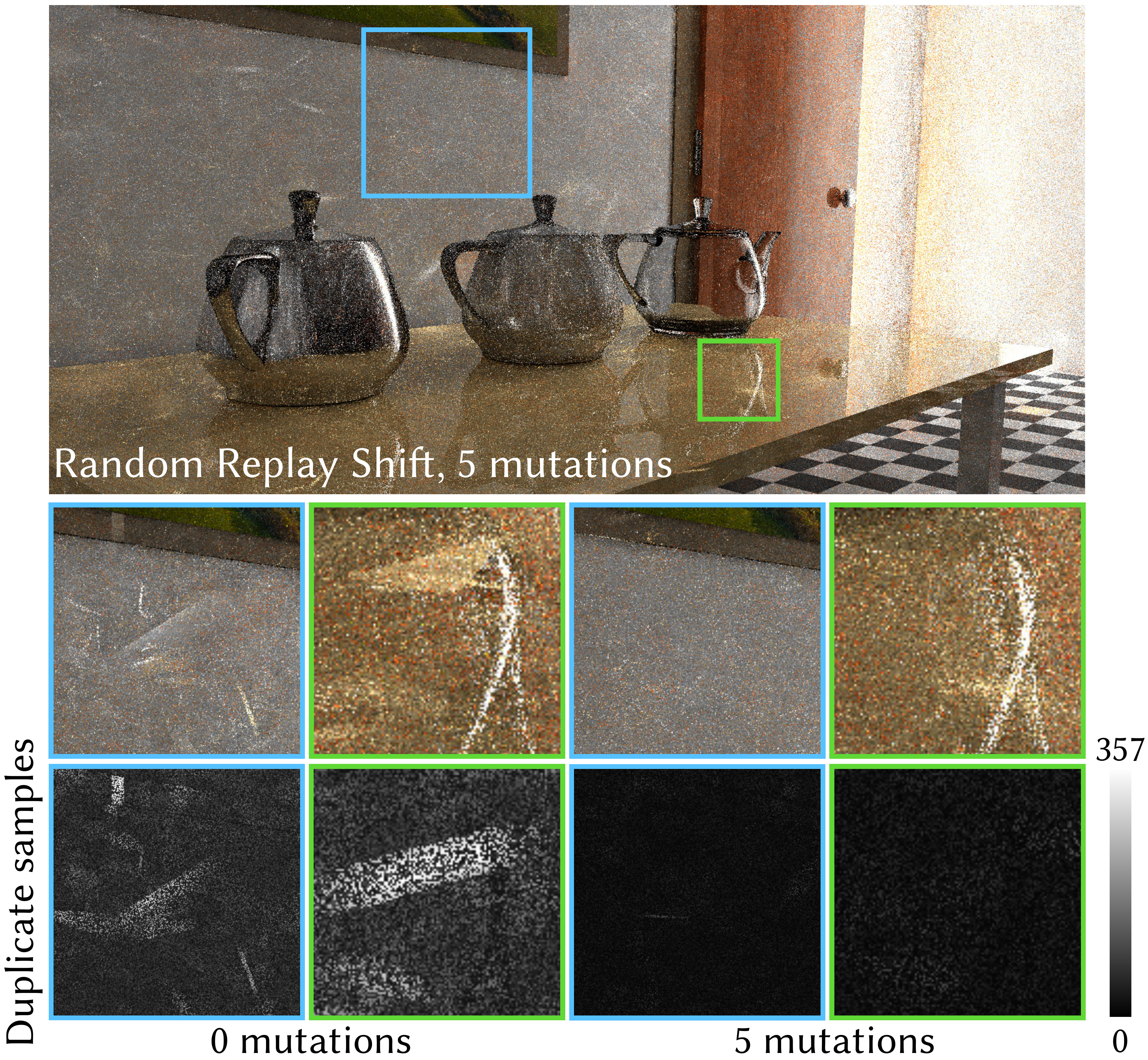}
    \caption{Mutations mitigate sample impoverishment in ReSTIR by diversifying the sample population. The bottom row visualizes duplicate samples in $20 \times 20$ pixel neighborhoods on the scene from \figref{SampleImpoverishment1}.\label{fig:ImprovedImpoverishment}}
\end{figure}

\begin{figure}[t]
    \centering
    \includegraphics[width=\columnwidth]{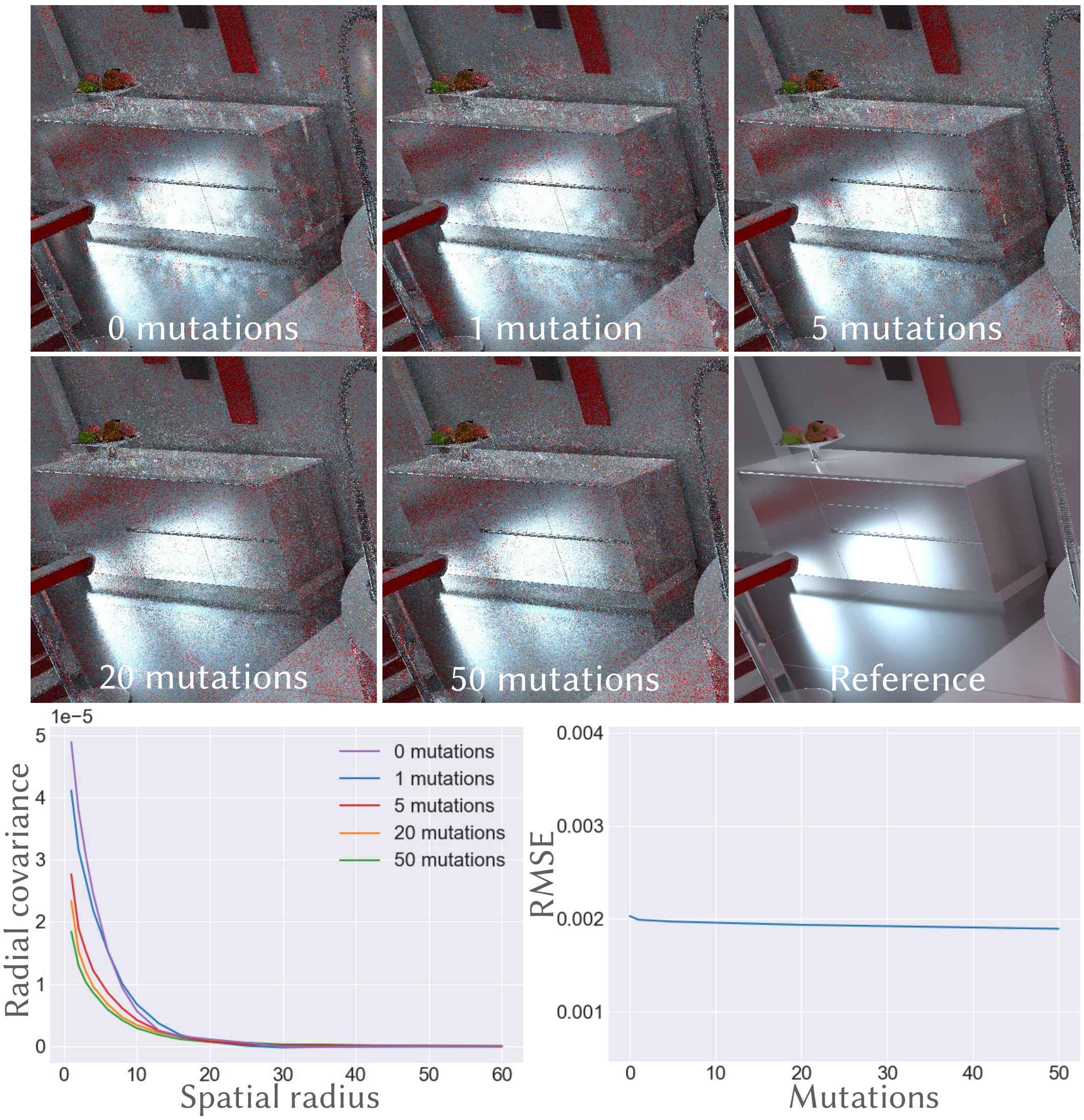}
    \caption{Increasing sample mutations reduces short range correlation artifacts produced by ReSTIR, with even 1-5 mutations providing noticeable improvements in image quality (measured in the bottom left using average radial covariance). Mutations typically have little impact on mean squared error (shown in the bottom right at equal spp with $M_{\text{cap}} = 50$) as we perturb samples only within each pixel.\label{fig:ImpactOnVariance}}
\end{figure}

\begin{figure}[t]
    \centering
    \includegraphics[width=\columnwidth]{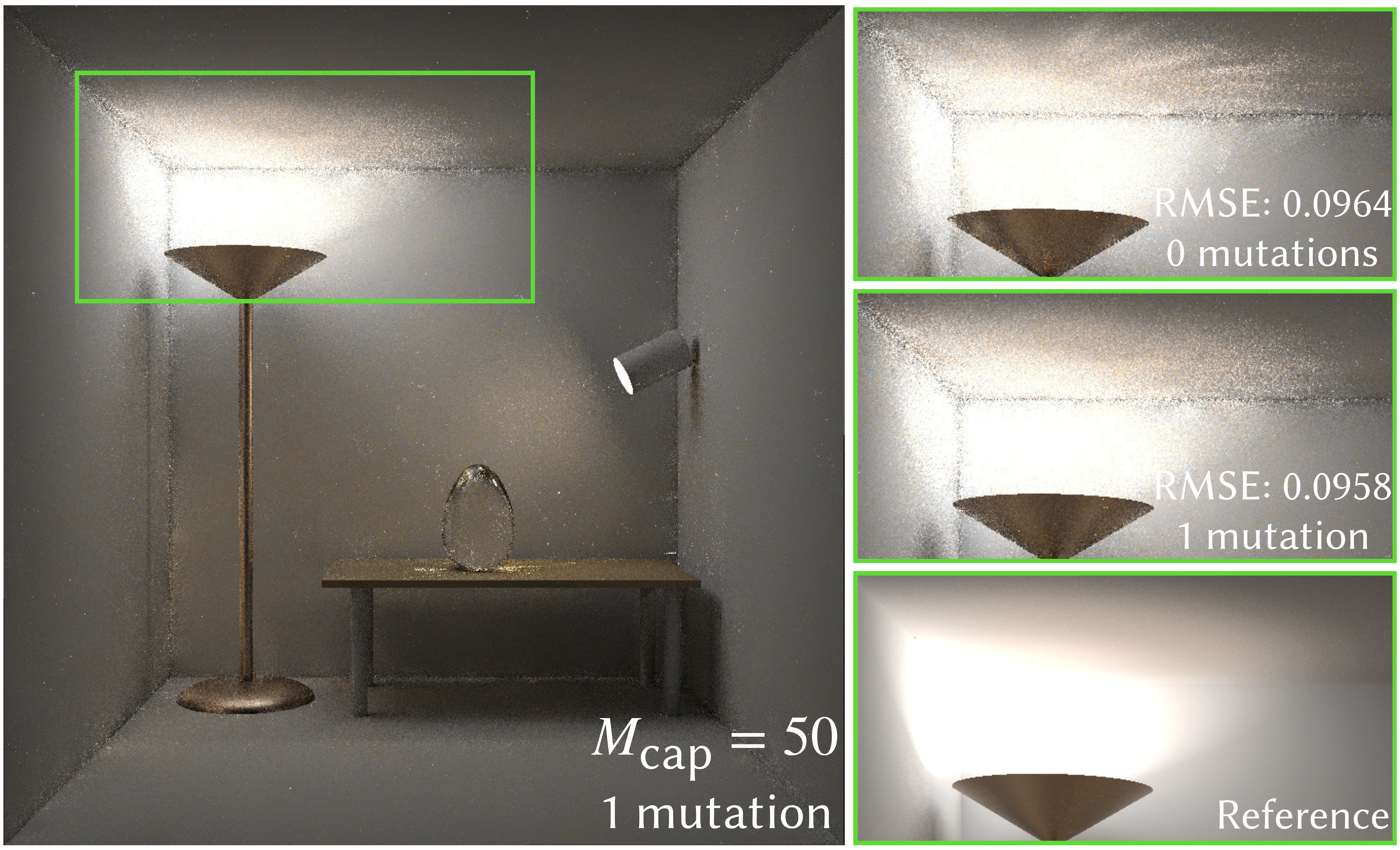}
    \caption{By reducing correlation artifacts, mutations allow use of larger $M_{\text{cap}}$ values in ReSTIR to trade noise for correlation, yielding lower error in scenes with difficult to sample light-carrying paths (see \figref{MCapArtifacts} for results with smaller $M_{\text{cap}}$ values).\label{fig:MCapAblation}}
\end{figure}

As shown in Figure ~\ref{fig:teaser} and Figures ~\ref{fig:EqualTimeComparison}--\ref{fig:MCapAblation}, short-range correlation artifacts are noticeably reduced in scenes with glossy materials and difficult lighting with just 1--5 mutations; further mutations have diminishing returns in improving image quality (\figref{ImpactOnVariance}). Mutation cost overhead is generally less than simply increasing sample count (\figref{EqualTimeComparison}), and recent denoisers \citep{Nvidia:2017:Optix} provide considerably better results with our decorrelated samples (see \figref{teaser}). \figref{ImprovedImpoverishment} shows mutations greatly reduce sample impoverishment, with fewer reservoirs sharing the exact same sample realizations.

Compared to standard path tracing, ReSTIR is much faster at achieving equal-error via correlated sampling for real-time direct \citep[Figure 8]{Bitterli:2020:ReSTIR} and global illumination \citep[Figure 13]{Lin:2022:GRIS}. Mutations however provide only marginal improvements in mean squared error in ReSTIR samplers (see Figures~\ref{fig:ImpactOnVariance} and ~\ref{fig:DifficultLightPathsMutations}). Akin to blue-noise dithering \citep{Georgiev:2016:BlueNoise, Heitz:2019:BlueNoise}, our image quality improves despite errors having similar magnitudes. The reason is mutating within a pixel leaves the sum of resampling weight unchanged in \eqref{ModifiedContributionWeightSubstituion}, and these weights ultimately control RIS estimator variance (\eqref{RIS2}). Mutations do slightly reduce variance, as they indirectly alter resampling weights of \emph{future} samples thanks to spatiotemporal reuse of the new, more diverse sample population; the supplementary document has more details. In \figref{MCapAblation} we also ablate $M_{\text{cap}}$ values to show the greater leeway our approach offers for this parameter, allowing use of larger values to trade noise for correlation.

\begin{table*}[t]
    \centering
    \includegraphics[width=\linewidth]{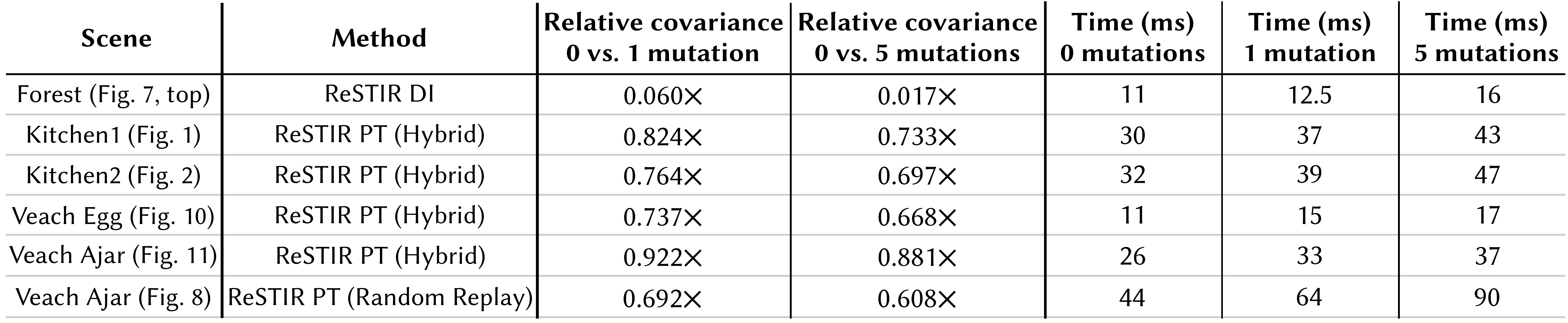}
    \caption{Reduction in covariance from mutations at 1 spp. We first average covariances over boxes of pixel radius 8, and then over the entire image. The Forest scene uses $M_{\text{cap}} = 20$; the rest use $M_{\text{cap}} = 50$. \label{tab:CovarianceNumbers}}
\end{table*}

Since ReSTIR often suffers from correlation artifacts, we quantify improvements in correlation by computing sample covariance between pixels, which naturally generalizes sample variance. This metric measures the joint variability of two random variables (\eg, whether error in two pixels varies similarly). For pixels $i$ and $j$ in image $I$, the sample covariance $c_{ij}$ between $i$ and $j$ is given by
\begin{equation}
    c_{ij} = \frac{1}{K-1} \sum_{k=1}^K \Big(I_{ki} - \bar{I}_{i}\Big)\Big(I_{kj} - \bar{I}_{j}\Big),
    \label{eq:SampleCovariance}
\end{equation}
where $K$ is the number of images used to estimate covariance (we use $K = 100$), and $\bar{I}$ is the average of $K$ images. To capture the joint variability of a pixel with its local neighborhood, in our experiments we average covariance estimates over boxes of a given radius centered at each pixel. We then further average over the entire image to get a single number. \figref{ImpactOnVariance} (bottom left) shows average radial covariance decreases with increasing spatial radius. This is expected as ReSTIR only reuses samples in local neighborhoods (so small-scale correlation artifacts are more pronounced); mutations reduce covariance in these short ranges. 

\tabref{CovarianceNumbers} lists the reduction in average covariance observed on our scenes, with pixel radius equal to 8. As correlations are typically localized, the reduction is even larger for the image insets in our figures compared to the results in \tabref{CovarianceNumbers}. Ineffective shift mappings in ReSTIR often result in increased correlations; mutations compensate for this shortcoming. For instance, mutations typically have fewer correlation artifacts to resolve with a hybrid shift in ReSTIR PT compared to, \eg, random replay (Figures~\ref{fig:DifficultLightPathsMutations} and ~\ref{fig:ImprovedImpoverishment} respectively), which highlights the benefit of using good shift mappings. In contrast, mutations provide greater covariance reduction in the Forest scene rendered with ReSTIR DI (\figref{EqualTimeComparison}, top), where higher covariance stems from vertex reconnections failing to preserve path contributions for low roughness surfaces.

\paragraph{Why mutations help?} The supplemental document details why mutations reduce covariance, simplifying down to the following, somewhat unintuitive, phenomenon: without mutations, covariance between pixels $i$ and $j$ stems from mismatches between input sample distributions and the target functions at $i$ and $j$ (Equation 16 in the supplemental). However, in the limit of infinite mutations, covariance is determined by samples' mismatch \emph{with their own pixel's} target function (Equation 10 in the supplemental) due to the ratio $\phat(x^0) / \phat(x^k)$ in the mutated contribution weight (\eqref{ModifiedContributionWeight}); this mismatch tends to be smaller. Though our analysis predicts that covariance does not vanish completely even with infinite mutations, our results show covariance is often reduced with just one mutation.

\section{Related Work}
\label{sec:RelatedWork}

Our method builds directly on the recent ReSTIR family of algorithms for real-time direct \citep{Bitterli:2020:ReSTIR} and global illumination \citep{Ouyang:2021:ReSTIRGI, Lin:2021:ReSTIRVol, Lin:2022:GRIS}. We augment spatiotemporal reservoir resampling in ReSTIR with sample mutations, and demonstrate the complementary strengths of resampling and mutations in this framework. In graphics, our approach is most closely related to Metropolis Light Transport (MLT) \citep{Veach:1997:MLT} and associated techniques \citep{Kelemen:2002:PSSMLT, Jakob:2012:Manifold, Lehtinen:2013:GDMLT, Hachisuka:2014:Multiplexed, Otsu:2018:Geometry, Cline:2005:ERPT, Lai:2007:PMCERPT, Lai:2009:PMCTemporal, Bashford:2021:EnsembleMLT}. In the broader Monte Carlo landscape, our approach belongs to the class of algorithms that jointly use resampling and mutations for sampling problems, such as Sequential Monte Carlo (SMC) \citep{Doucet:2001:SMC} and Population Monte Carlo (PMC) \citep{Cappe:2004:PMC}. We discuss the relation to MLT, SMC and PMC in more detail next; \tabref{RelatedWorkTable} provides a summary. We refer the reader to \citet[Section 7]{Bitterli:2020:ReSTIR} and \citet[Section 9.3]{Lin:2022:GRIS} for comparisons between ReSTIR and other rendering algorithms that exploit path reuse and spatial correlations.

\begin{table*}
    \centering
    \includegraphics[width=\linewidth]{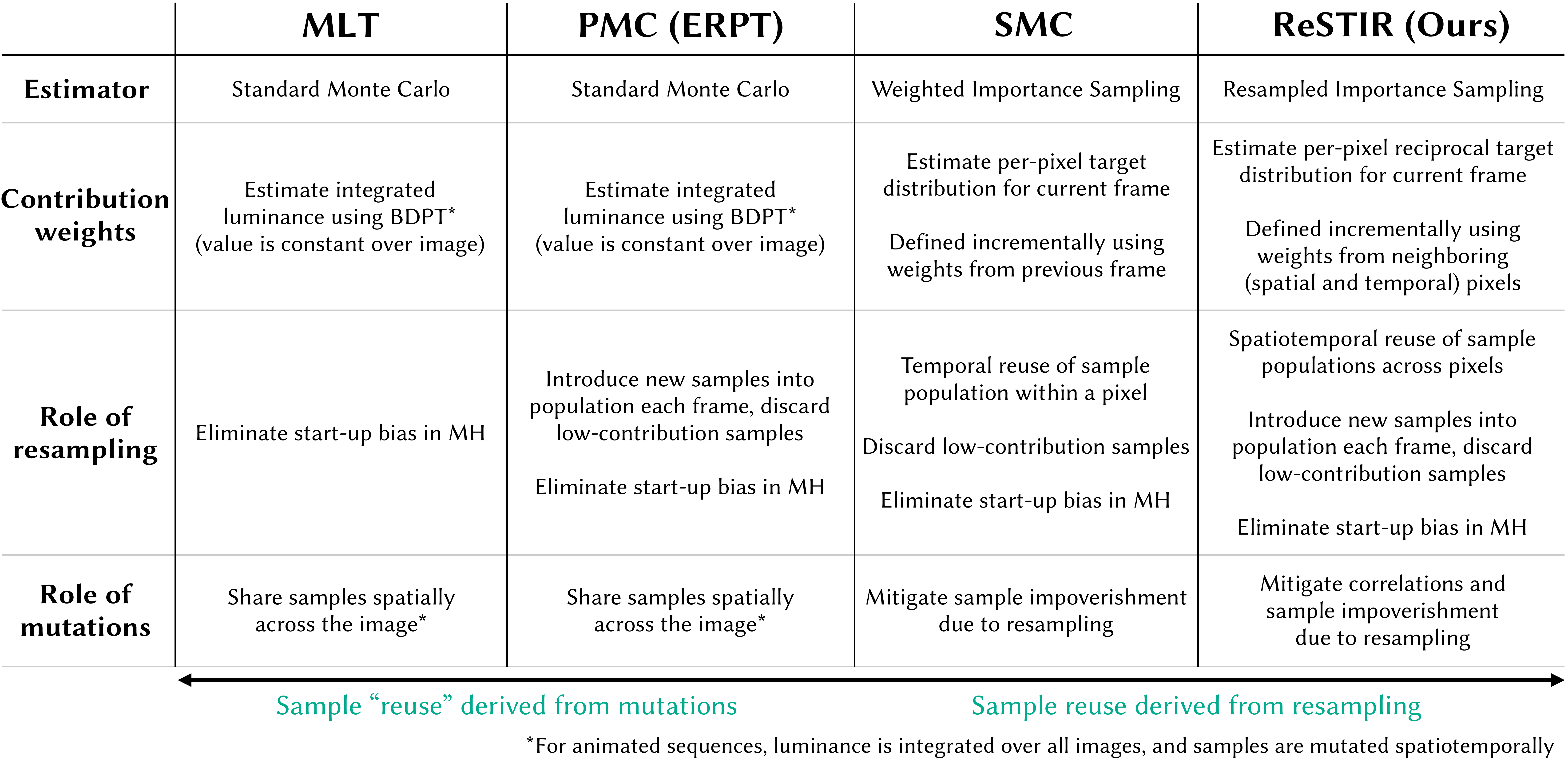}
    \caption{Overview of the role of resampling and mutations in MLT, PMC, SMC and ReSTIR.}
    \label{tab:RelatedWorkTable}
\end{table*}

\paragraph{Metropolis Light Transport} MLT uses statistically correlated samples generated by Metropolis--Hastings to solve the rendering equation. Unlike algorithms using independent samples, MLT is effective at finding difficult light paths by locally exploring the path space. It reuses samples by mutating high-contribution paths over the image. Algorithmically, our method resembles MLT in various ways. Both techniques require secondary estimators, respectively RIS and bidirectional path tracing (BDPT) \citep{Lafortune:1993:Bidirectional, Veach:1995:Bidirectional}, to normalize the MH target function. Samples used by these estimators are resampled into a smaller set to initialize MH (our \secref{Method} and \citet[Chapter 11.3.1]{Veach:1998:Robust}),
and contributions of mutated samples are effectively weighted by the same weights (\eqref{ModifiedContributionWeightSubstituion}) to remain unbiased (our \appref{StartupBiasElimination} and \citet[Appendix 11.A]{Veach:1998:Robust}).

\begin{figure}[t]
    \centering
    \includegraphics[width=\columnwidth]{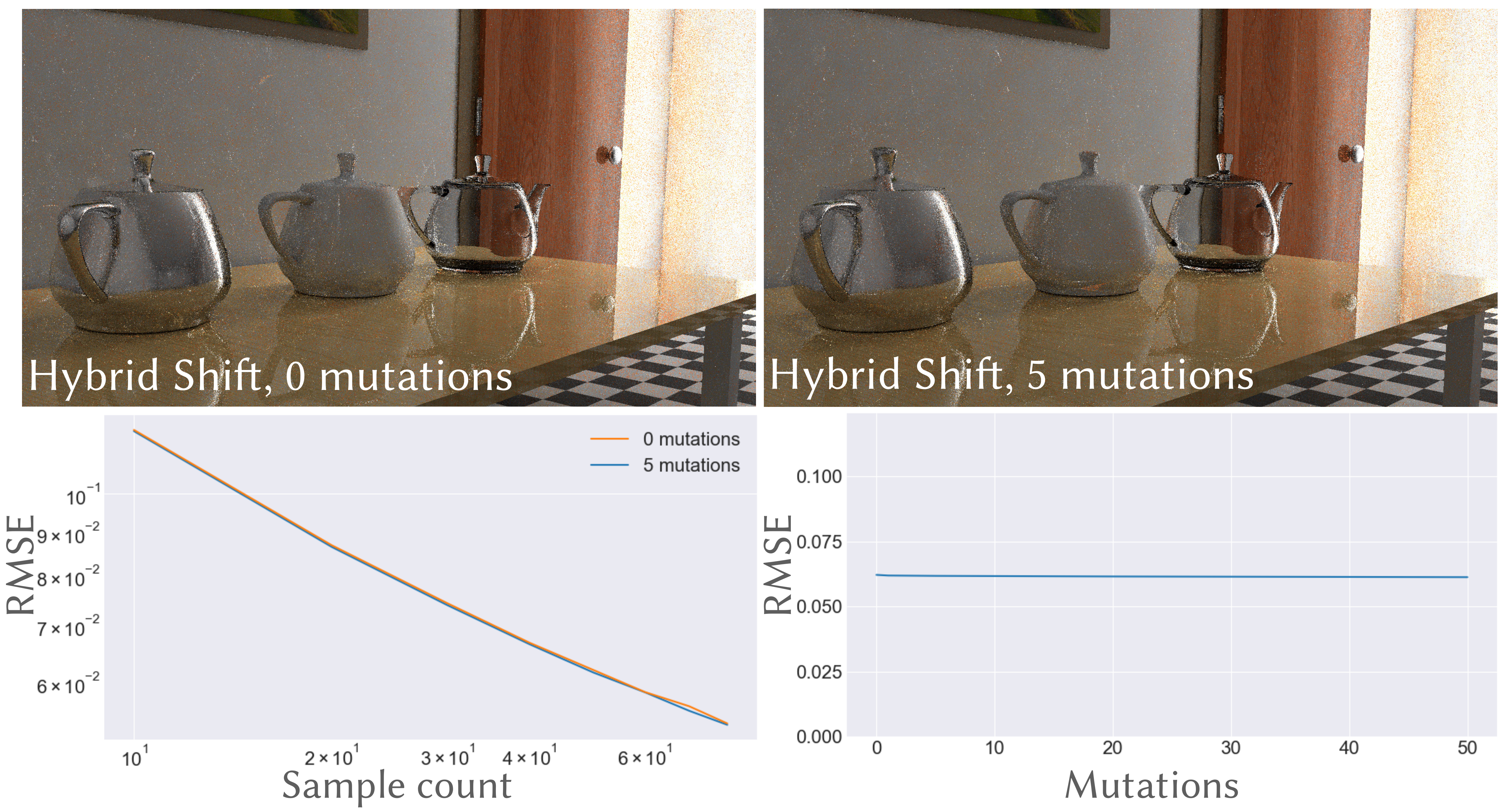}
    \caption{Mutations do not reduce mean squared error in the Veach Ajar scene rendered using ReSTIR PT with the hybrid shift and $M_{\text{cap}} = 50$. This suggests that in contrast to Metropolis Light Transport, resampling (and not mutations) finds important light-carrying paths in ReSTIR. Compared to the random replay shift in \figref{ImprovedImpoverishment}, resampling with the superior hybrid shift does not introduce large correlation artifacts in this scene.\label{fig:DifficultLightPathsMutations}}
\end{figure}

The crucial difference between our work and MLT lies in how samples are reused across pixels. MLT latches onto high-contribution paths and mutates them over the entire image while just resampling to eliminate start-up bias. Thus, MLT results often contain correlation artifacts caused by mutations, applying MH to both find important samples and redistribute them between pixels. In contrast, ReSTIR derives \emph{spatiotemporal} reuse from resampling; in this paper, we mutate samples
within each pixel
to \emph{mitigate} correlations and sample impoverishment from spatiotemporal resampling. As a result, our method does not require numerous MH iterations, as the primary purpose of mutations is not finding important paths (\figref{DifficultLightPathsMutations}). Further, our approach suits real-time rendering as it integrates seamlessly into ReSTIR. MLT can be adapted to mutate temporally, but unlike our work, the entire animated sequence must be available in advance \citep{Van:2017:MLTTemporal}.

Several features have recently been added to MLT, including sample stratification \citep{Cline:2005:ERPT}, MIS \citep{Hachisuka:2014:Multiplexed} and enhanced mutation strategies \citep{Jakob:2012:Manifold, Bitterli:2017:Reversible, Otsu:2018:Geometry, Kaplanyan:2014:HalfVectorMLT}. Though we mostly employ simple PSS-style mutations \citep{Kelemen:2002:PSSMLT}, many of these improvements can also be incorporated into our approach.

\begin{wrapfigure}[10]{r}{0.24\textwidth}
  \centering
    \vspace{-0.25em}\hspace{-2.48em}\includegraphics[width=0.28\textwidth]{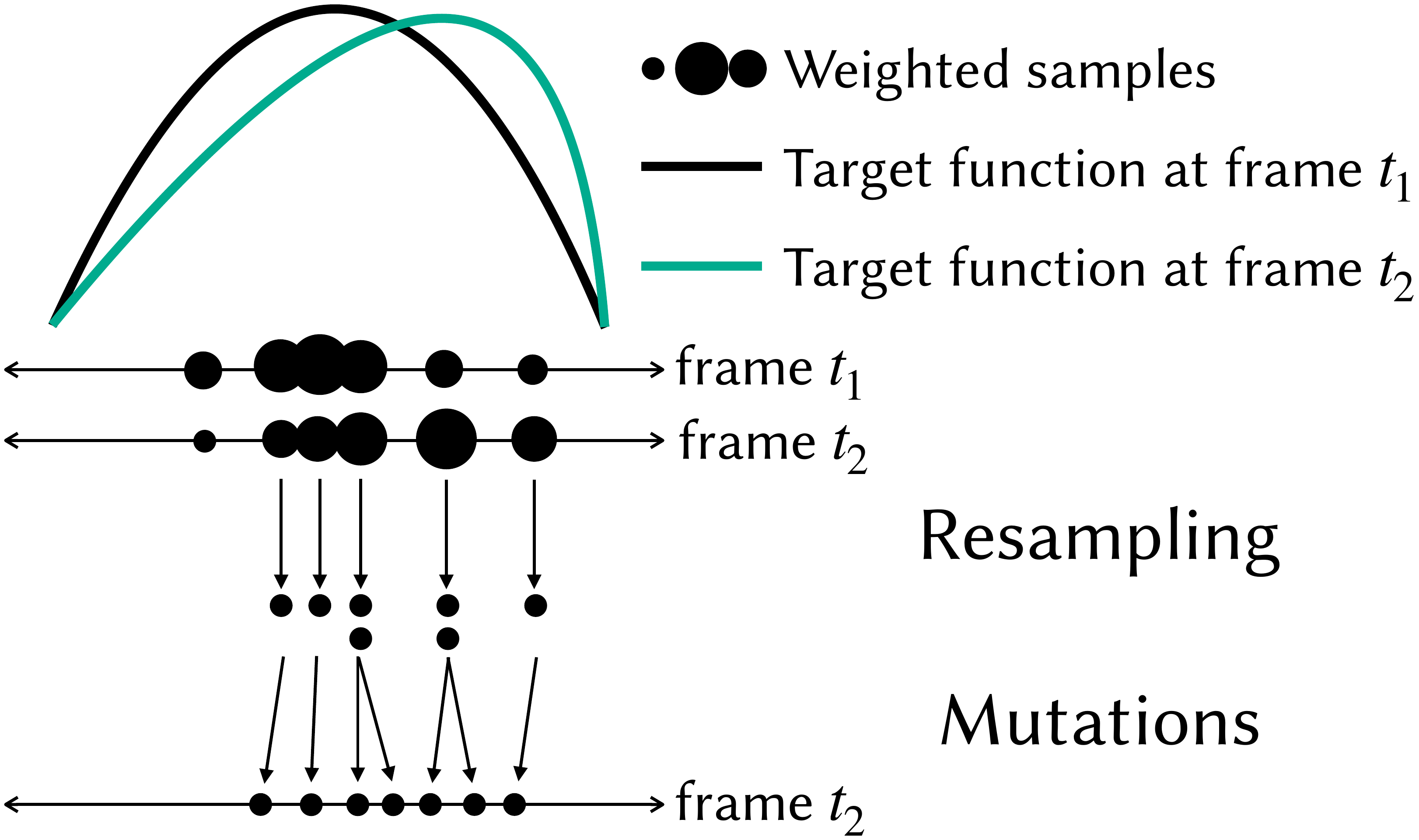}
  \label{fig:smc-pipeline}
\end{wrapfigure}
\paragraph{Sequential Monte Carlo} SMC is a family of Monte Carlo methods used for filtering and tracking in Bayesian inference and signal processing \citep{Doucet:2001:SMC}. As shown in the inset, the goal is maintaining a population of weighted samples distributed roughly proportional to an evolving target distribution (with unknown normalization factor). Sample weights are adjusted every iteration to reflect each sample's importance to the most recent distribution. Resampling discards samples with low weights and duplicates those with high weights. Mutations ensure the population does not contain identical samples. Unlike ReSTIR, which uses RIS, SMC methods use \emph{weighted importance sampling (WIS)} to estimate correlated integrals
in a chained fashion, \ie, the current step's sample weights and normalization factors are defined incrementally based on corresponding quantities from earlier steps \citep{Del:2006:SMCSamplers}. This allows temporally reusing samples for estimation, instead of generating new samples every frame.

SMC methods have found limited use in rendering; \citet{Ghosh:2006:SMCDI} sample a sequence of per-pixel target functions for direct illumination of dynamic environment maps. Unlike ReSTIR, which derives its samples from spatiotemporally neighboring pixels, \citet{Ghosh:2006:SMCDI} instead maintain a fixed sample population per pixel that is resampled and mutated to be updated for each frame. Large populations are needed for effective importance sampling, as high-contribution samples are not shared between pixels; in contrast, ReSTIR often stores just a single sample per reservoir. SMC methods likely require MIS weights and shift mappings (like ReSTIR) to resolve bias and correctly derive effective spatiotemporal reuse from neighbors. Similar to our work, mutations mitigate sample impoverishment but do not provide reuse.

\paragraph{Population Monte Carlo} PMC methods also couple resampling and mutations to distribute weighted samples in proportion to a sequence of target functions \citep{Cappe:2004:PMC}. The main added feature is they sample using parametric mixture models with simple source PDFs. Mixture probabilities are tuned for each target function using previously generated samples and their importance. 

In rendering, the PMC framework has been used for direct lighting \citep{Fan:2007:PMCDI, Lai:2015:PMCDI}, global illumination \citep{Lai:2007:PMCPT, Lai:2007:PMCERPT} and animation \citep{Lai:2009:PMCTemporal}. \citeauthor{Lai:2009:PMCTemporal}'s [\citeyear{Lai:2009:PMCTemporal}] work is most relevant to ours: they derive sample reuse by mutating samples spatially and temporally across the image plane using Energy Redistribution Path Tracing (ERPT) \citep{Cline:2005:ERPT}. Resampling serves to select high-contribution samples while discarding  those with small weights; it is also used to refresh the sample population (much like initial resampling in ReSTIR) and eliminate start-up bias from mutations. Unlike our method, they require knowing animated sequences in advance, precluding most real-time applications.

\section{Limitations and Future Work}
\label{sec:LimitationsAndFutureWork}

In this paper, we provide an unbiased mechanism leveraging MCMC mutations to diversify ReSTIR's sample population.
Often, just a single mutation per pixel effectively mitigates correlation artifacts in glossy scenes with complex lighting.
However, as in most MCMC schemes, we cannot accurately predict the number of Metropolis--Hastings iterations needed to reduce correlations below a given threshold.
Beyond the analysis in the supplemental document, further investigation is also needed to understand how mutations address sample impoverishment in ReSTIR---not just in terms of the number of duplicate samples (\figref{ImprovedImpoverishment}), but also the discrepancy characteristics of the resulting sample population.

Mutating inside ReSTIR has a non-negligible run-time overhead.
Though we demonstrate improvements on an equal-time covariance metric with simple mutation strategies in both ReSTIR DI and ReSTIR PT (\figref{EqualTimeComparison} and \tabref{CovarianceNumbers}), more sophisticated mutations \citep{Jakob:2012:Manifold, Bitterli:2017:Reversible, Otsu:2018:Geometry, Kaplanyan:2014:HalfVectorMLT} could provide further gains.
Our decision to mutate only after temporal (but not spatial) resampling is also informed in part by run-time considerations.
As mentioned in \secref{Method}, applying mutations selectively (\ie, not at each pixel every frame) based on \eg, local correlation heuristics could improve performance.
As both mutations and ReSTIR's initial path candidates serve to rejuvenate the sample population, it may be interesting to carefully balance the costs of per-pixel mutations versus new path candidates. 

Our proposed sample mutations reduce correlation between nearby pixels, leading to an error distribution (likely) closer to white noise.
But blue noise error distributions are often superior with respect to human perception~\citep{Mitchell:1987:antialiased}; perhaps our mutations could change to more directly optimize for blue noise characteristics.
For example, when deciding mutation acceptance, we might consider both the target function and the neighboring pixel samples, preferring mutations that introduce differing sample values.
A further improvement might apply the insights of \citet{Heitz:2019:BlueNoise} to optimize the image-space distribution of error rather than solely considering the sample values.

Like Metropolis Light Transport, mutating samples across pixels potentially unlocks further amortization by sharing samples over the entire image (\eg, using the \emph{expected values} technique \citep[Section 11.5]{Veach:1998:Robust}).
We leave such "cross-pixel" mutations to future work as they require adjusting a mutated sample's contribution weight (\eqref{ModifiedContributionWeight}) to account for varying integration domains.

More generally, by augmenting ReSTIR with mutations, our work establishes a closer correspondence between the RIS-based resampling techniques developed in graphics, and those in the broader statistics literature such as SMC and PMC. In particular, our approach stands to benefit from techniques such as annealed importance sampling \citep{Neal:2001:AnnealedIS} used in SMC to reduce variance in the resampling weights \citep[Section 4]{Ghosh:2006:SMCDI}, as well as from adaptation strategies for mutation kernels developed in PMC to increase acceptance rates \citep[Section 4.2]{Lai:2007:PMCERPT}. Moreover, as in these fields, mutations in ReSTIR open the door not just to artifact-free integration
(of the rendering equation),
but also to tracking and filtering problems---for instance using well-distributed sample populations generated by our approach as training data for path guiding.

\bibliographystyle{ACM-Reference-Format}
\bibliography{RestirMCMC}


\begin{thebibliography}{54}


\ifx \showCODEN    \undefined \def \showCODEN     #1{\unskip}     \fi
\ifx \showDOI      \undefined \def \showDOI       #1{#1}\fi
\ifx \showISBNx    \undefined \def \showISBNx     #1{\unskip}     \fi
\ifx \showISBNxiii \undefined \def \showISBNxiii  #1{\unskip}     \fi
\ifx \showISSN     \undefined \def \showISSN      #1{\unskip}     \fi
\ifx \showLCCN     \undefined \def \showLCCN      #1{\unskip}     \fi
\ifx \shownote     \undefined \def \shownote      #1{#1}          \fi
\ifx \showarticletitle \undefined \def \showarticletitle #1{#1}   \fi
\ifx \showURL      \undefined \def \showURL       {\relax}        \fi
\providecommand\bibfield[2]{#2}
\providecommand\bibinfo[2]{#2}
\providecommand\natexlab[1]{#1}
\providecommand\showeprint[2][]{arXiv:#2}

\bibitem[Bashford-Rogers et~al\mbox{.}(2021)]%
        {Bashford:2021:EnsembleMLT}
\bibfield{author}{\bibinfo{person}{Thomas Bashford-Rogers},
  \bibinfo{person}{Lu{\'\i}s~Paulo Santos}, \bibinfo{person}{Demetris
  Marnerides}, {and} \bibinfo{person}{Kurt Debattista}.}
  \bibinfo{year}{2021}\natexlab{}.
\newblock \showarticletitle{Ensemble Metropolis Light Transport}.
\newblock \bibinfo{journal}{\emph{ACM Transactions on Graphics (TOG)}}
  \bibinfo{volume}{41}, \bibinfo{number}{1} (\bibinfo{year}{2021}),
  \bibinfo{pages}{1--15}.
\newblock


\bibitem[Bitterli(2022)]%
        {Bitterli:2022:Correlations}
\bibfield{author}{\bibinfo{person}{Benedikt Bitterli}.}
  \bibinfo{year}{2022}\natexlab{}.
\newblock \bibinfo{booktitle}{\emph{Correlations and Reuse for Fast and
  Accurate Physically Based Light Transport}}. Vol.~\bibinfo{volume}{77}.
\newblock \bibinfo{publisher}{Dartmouth College Ph.D Dissertations}.
\newblock
\urldef\tempurl%
\url{https://digitalcommons.dartmouth.edu/dissertations/77}
\showURL{%
\tempurl}


\bibitem[Bitterli et~al\mbox{.}(2017)]%
        {Bitterli:2017:Reversible}
\bibfield{author}{\bibinfo{person}{Benedikt Bitterli}, \bibinfo{person}{Wenzel
  Jakob}, \bibinfo{person}{Jan Nov{\'a}k}, {and} \bibinfo{person}{Wojciech
  Jarosz}.} \bibinfo{year}{2017}\natexlab{}.
\newblock \showarticletitle{Reversible jump Metropolis light transport using
  inverse mappings}.
\newblock \bibinfo{journal}{\emph{ACM Transactions on Graphics (TOG)}}
  \bibinfo{volume}{37}, \bibinfo{number}{1} (\bibinfo{year}{2017}),
  \bibinfo{pages}{1--12}.
\newblock


\bibitem[Bitterli et~al\mbox{.}(2020)]%
        {Bitterli:2020:ReSTIR}
\bibfield{author}{\bibinfo{person}{Benedikt Bitterli}, \bibinfo{person}{Chris
  Wyman}, \bibinfo{person}{Matt Pharr}, \bibinfo{person}{Peter Shirley},
  \bibinfo{person}{Aaron Lefohn}, {and} \bibinfo{person}{Wojciech Jarosz}.}
  \bibinfo{year}{2020}\natexlab{}.
\newblock \showarticletitle{Spatiotemporal reservoir resampling for real-time
  ray tracing with dynamic direct lighting}.
\newblock \bibinfo{journal}{\emph{ACM Transactions on Graphics (TOG)}}
  \bibinfo{volume}{39}, \bibinfo{number}{4} (\bibinfo{year}{2020}),
  \bibinfo{pages}{148--1}.
\newblock


\bibitem[Capp{\'e} et~al\mbox{.}(2004)]%
        {Cappe:2004:PMC}
\bibfield{author}{\bibinfo{person}{Olivier Capp{\'e}}, \bibinfo{person}{Arnaud
  Guillin}, \bibinfo{person}{Jean-Michel Marin}, {and}
  \bibinfo{person}{Christian~P Robert}.} \bibinfo{year}{2004}\natexlab{}.
\newblock \showarticletitle{Population {M}onte {C}arlo}.
\newblock \bibinfo{journal}{\emph{Journal of Computational and Graphical
  Statistics}} \bibinfo{volume}{13}, \bibinfo{number}{4}
  (\bibinfo{year}{2004}), \bibinfo{pages}{907--929}.
\newblock


\bibitem[Chaitanya et~al\mbox{.}(2017)]%
        {Chaitanya:2017:Denoising}
\bibfield{author}{\bibinfo{person}{Chakravarty R~Alla Chaitanya},
  \bibinfo{person}{Anton~S Kaplanyan}, \bibinfo{person}{Christoph Schied},
  \bibinfo{person}{Marco Salvi}, \bibinfo{person}{Aaron Lefohn},
  \bibinfo{person}{Derek Nowrouzezahrai}, {and} \bibinfo{person}{Timo Aila}.}
  \bibinfo{year}{2017}\natexlab{}.
\newblock \showarticletitle{Interactive reconstruction of Monte Carlo image
  sequences using a recurrent denoising autoencoder}.
\newblock \bibinfo{journal}{\emph{ACM Transactions on Graphics (TOG)}}
  \bibinfo{volume}{36}, \bibinfo{number}{4} (\bibinfo{year}{2017}),
  \bibinfo{pages}{1--12}.
\newblock


\bibitem[Chao(1982)]%
        {Chao:1982:WRS}
\bibfield{author}{\bibinfo{person}{Min-Te Chao}.}
  \bibinfo{year}{1982}\natexlab{}.
\newblock \showarticletitle{A general purpose unequal probability sampling
  plan}.
\newblock \bibinfo{journal}{\emph{Biometrika}} \bibinfo{volume}{69},
  \bibinfo{number}{3} (\bibinfo{year}{1982}), \bibinfo{pages}{653--656}.
\newblock


\bibitem[Cline et~al\mbox{.}(2005)]%
        {Cline:2005:ERPT}
\bibfield{author}{\bibinfo{person}{David Cline}, \bibinfo{person}{Justin
  Talbot}, {and} \bibinfo{person}{Parris Egbert}.}
  \bibinfo{year}{2005}\natexlab{}.
\newblock \showarticletitle{Energy redistribution path tracing}.
\newblock \bibinfo{journal}{\emph{ACM Transactions on Graphics (TOG)}}
  \bibinfo{volume}{24}, \bibinfo{number}{3} (\bibinfo{year}{2005}),
  \bibinfo{pages}{1186--1195}.
\newblock


\bibitem[Dachsbacher et~al\mbox{.}(2014)]%
        {Dachsbacher:2014:ManyLight}
\bibfield{author}{\bibinfo{person}{Carsten Dachsbacher},
  \bibinfo{person}{Jaroslav K{\v{r}}iv{\'a}nek}, \bibinfo{person}{Milo{\v{s}}
  Ha{\v{s}}an}, \bibinfo{person}{Adam Arbree}, \bibinfo{person}{Bruce Walter},
  {and} \bibinfo{person}{Jan Nov{\'a}k}.} \bibinfo{year}{2014}\natexlab{}.
\newblock \showarticletitle{Scalable realistic rendering with many-light
  methods}. In \bibinfo{booktitle}{\emph{Computer Graphics Forum}},
  Vol.~\bibinfo{volume}{33}. Wiley Online Library, \bibinfo{pages}{88--104}.
\newblock


\bibitem[Del~Moral et~al\mbox{.}(2006)]%
        {Del:2006:SMCSamplers}
\bibfield{author}{\bibinfo{person}{Pierre Del~Moral}, \bibinfo{person}{Arnaud
  Doucet}, {and} \bibinfo{person}{Ajay Jasra}.}
  \bibinfo{year}{2006}\natexlab{}.
\newblock \showarticletitle{Sequential {M}onte {C}arlo samplers}.
\newblock \bibinfo{journal}{\emph{Journal of the Royal Statistical Society:
  Series B (Statistical Methodology)}} \bibinfo{volume}{68},
  \bibinfo{number}{3} (\bibinfo{year}{2006}), \bibinfo{pages}{411--436}.
\newblock


\bibitem[Doucet et~al\mbox{.}(2001)]%
        {Doucet:2001:SMC}
\bibfield{author}{\bibinfo{person}{Arnaud Doucet}, \bibinfo{person}{Nando
  De~Freitas}, \bibinfo{person}{Neil~James Gordon}, {et~al\mbox{.}}}
  \bibinfo{year}{2001}\natexlab{}.
\newblock \bibinfo{booktitle}{\emph{Sequential Monte Carlo Methods in
  Practice}}. Vol.~\bibinfo{volume}{1}.
\newblock \bibinfo{publisher}{Springer}.
\newblock


\bibitem[Fan et~al\mbox{.}(2007)]%
        {Fan:2007:PMCDI}
\bibfield{author}{\bibinfo{person}{ShaoHua Fan}, \bibinfo{person}{Yu-Chi Lai},
  \bibinfo{person}{Stephen Chenney}, {and} \bibinfo{person}{Charles Dyer}.}
  \bibinfo{year}{2007}\natexlab{}.
\newblock \bibinfo{booktitle}{\emph{Population Monte Carlo Samplers for
  Rendering}}.
\newblock \bibinfo{type}{{T}echnical {R}eport}.
  \bibinfo{institution}{University of Wisconsin-Madison Department of Computer
  Sciences}.
\newblock


\bibitem[Georgiev and Fajardo(2016)]%
        {Georgiev:2016:BlueNoise}
\bibfield{author}{\bibinfo{person}{Iliyan Georgiev} {and}
  \bibinfo{person}{Marcos Fajardo}.} \bibinfo{year}{2016}\natexlab{}.
\newblock \showarticletitle{Blue-noise dithered sampling}.
\newblock In \bibinfo{booktitle}{\emph{ACM SIGGRAPH 2016 Talks}}.
  \bibinfo{pages}{1--1}.
\newblock


\bibitem[Ghosh et~al\mbox{.}(2006)]%
        {Ghosh:2006:SMCDI}
\bibfield{author}{\bibinfo{person}{Abhijeet Ghosh}, \bibinfo{person}{Arnaud
  Doucet}, {and} \bibinfo{person}{Wolfgang Heidrich}.}
  \bibinfo{year}{2006}\natexlab{}.
\newblock \showarticletitle{Sequential Sampling for Dynamic Environment Map
  Illumination.}. In \bibinfo{booktitle}{\emph{Rendering Techniques}}.
  \bibinfo{pages}{115--126}.
\newblock


\bibitem[Hachisuka and Jensen(2009)]%
        {Hachisuka:2009:SPPM}
\bibfield{author}{\bibinfo{person}{Toshiya Hachisuka} {and}
  \bibinfo{person}{Henrik~Wann Jensen}.} \bibinfo{year}{2009}\natexlab{}.
\newblock \showarticletitle{Stochastic progressive photon mapping}.
\newblock In \bibinfo{booktitle}{\emph{ACM SIGGRAPH Asia 2009 papers}}.
  \bibinfo{pages}{1--8}.
\newblock


\bibitem[Hachisuka et~al\mbox{.}(2014)]%
        {Hachisuka:2014:Multiplexed}
\bibfield{author}{\bibinfo{person}{Toshiya Hachisuka}, \bibinfo{person}{Anton~S
  Kaplanyan}, {and} \bibinfo{person}{Carsten Dachsbacher}.}
  \bibinfo{year}{2014}\natexlab{}.
\newblock \showarticletitle{Multiplexed {M}etropolis light transport}.
\newblock \bibinfo{journal}{\emph{ACM Transactions on Graphics (TOG)}}
  \bibinfo{volume}{33}, \bibinfo{number}{4} (\bibinfo{year}{2014}),
  \bibinfo{pages}{1--10}.
\newblock


\bibitem[Hastings(1970)]%
        {Hastings:1970:MH}
\bibfield{author}{\bibinfo{person}{W~Keith Hastings}.}
  \bibinfo{year}{1970}\natexlab{}.
\newblock \showarticletitle{Monte Carlo sampling methods using Markov chains
  and their applications}.
\newblock  (\bibinfo{year}{1970}).
\newblock


\bibitem[Heitz and Belcour(2019)]%
        {Heitz:2019:BlueNoise}
\bibfield{author}{\bibinfo{person}{Eric Heitz} {and} \bibinfo{person}{Laurent
  Belcour}.} \bibinfo{year}{2019}\natexlab{}.
\newblock \showarticletitle{Distributing Monte Carlo errors as a blue noise in
  screen space by permuting pixel seeds between frames}. In
  \bibinfo{booktitle}{\emph{Computer Graphics Forum}},
  Vol.~\bibinfo{volume}{38}. Wiley Online Library, \bibinfo{pages}{149--158}.
\newblock


\bibitem[Hua et~al\mbox{.}(2019)]%
        {Hua:2019:GDPTSurvey}
\bibfield{author}{\bibinfo{person}{Binh-Son Hua}, \bibinfo{person}{Adrien
  Gruson}, \bibinfo{person}{Victor Petitjean}, \bibinfo{person}{Matthias
  Zwicker}, \bibinfo{person}{Derek Nowrouzezahrai}, \bibinfo{person}{Elmar
  Eisemann}, {and} \bibinfo{person}{Toshiya Hachisuka}.}
  \bibinfo{year}{2019}\natexlab{}.
\newblock \showarticletitle{A Survey on Gradient-Domain Rendering}. In
  \bibinfo{booktitle}{\emph{Computer Graphics Forum}},
  Vol.~\bibinfo{volume}{38}. Wiley Online Library, \bibinfo{pages}{455--472}.
\newblock


\bibitem[Jakob and Marschner(2012)]%
        {Jakob:2012:Manifold}
\bibfield{author}{\bibinfo{person}{Wenzel Jakob} {and} \bibinfo{person}{Steve
  Marschner}.} \bibinfo{year}{2012}\natexlab{}.
\newblock \showarticletitle{Manifold exploration: A {M}arkov chain {M}onte
  {C}arlo technique for rendering scenes with difficult specular transport}.
\newblock \bibinfo{journal}{\emph{ACM Transactions on Graphics (TOG)}}
  \bibinfo{volume}{31}, \bibinfo{number}{4} (\bibinfo{year}{2012}),
  \bibinfo{pages}{1--13}.
\newblock


\bibitem[Jensen(1996)]%
        {Jensen:1996:Photon}
\bibfield{author}{\bibinfo{person}{Henrik~Wann Jensen}.}
  \bibinfo{year}{1996}\natexlab{}.
\newblock \showarticletitle{Global illumination using photon maps}. In
  \bibinfo{booktitle}{\emph{Eurographics workshop on Rendering techniques}}.
  Springer, \bibinfo{pages}{21--30}.
\newblock


\bibitem[Kajiya(1986)]%
        {Kajiya:1986:Rendering}
\bibfield{author}{\bibinfo{person}{James~T Kajiya}.}
  \bibinfo{year}{1986}\natexlab{}.
\newblock \showarticletitle{The rendering equation}. In
  \bibinfo{booktitle}{\emph{Proceedings of The 13th Annual Conference on
  Computer Graphics and Interactive Techniques}}. \bibinfo{pages}{143--150}.
\newblock


\bibitem[Kallweit et~al\mbox{.}(2022)]%
        {Kallweit:2022:Falcor}
\bibfield{author}{\bibinfo{person}{Simon Kallweit}, \bibinfo{person}{Petrik
  Clarberg}, \bibinfo{person}{Craig Kolb}, \bibinfo{person}{Tom{'a}{\v s}
  Davidovi{\v c}}, \bibinfo{person}{Kai-Hwa Yao}, \bibinfo{person}{Theresa
  Foley}, \bibinfo{person}{Yong He}, \bibinfo{person}{Lifan Wu},
  \bibinfo{person}{Lucy Chen}, \bibinfo{person}{Tomas Akenine-M{\"o}ller},
  \bibinfo{person}{Chris Wyman}, \bibinfo{person}{Cyril Crassin}, {and}
  \bibinfo{person}{Nir Benty}.} \bibinfo{year}{2022}\natexlab{}.
\newblock \bibinfo{title}{The {Falcor} Rendering Framework}.
\newblock
\newblock
\urldef\tempurl%
\url{https://github.com/NVIDIAGameWorks/Falcor}
\showURL{%
\tempurl}
\newblock
\shownote{\url{https://github.com/NVIDIAGameWorks/Falcor}}.


\bibitem[Kaplanyan et~al\mbox{.}(2014)]%
        {Kaplanyan:2014:HalfVectorMLT}
\bibfield{author}{\bibinfo{person}{Anton~S Kaplanyan},
  \bibinfo{person}{Johannes Hanika}, {and} \bibinfo{person}{Carsten
  Dachsbacher}.} \bibinfo{year}{2014}\natexlab{}.
\newblock \showarticletitle{The natural-constraint representation of the path
  space for efficient light transport simulation}.
\newblock \bibinfo{journal}{\emph{ACM Transactions on Graphics (TOG)}}
  \bibinfo{volume}{33}, \bibinfo{number}{4} (\bibinfo{year}{2014}),
  \bibinfo{pages}{1--13}.
\newblock


\bibitem[Kelemen et~al\mbox{.}(2002)]%
        {Kelemen:2002:PSSMLT}
\bibfield{author}{\bibinfo{person}{Csaba Kelemen},
  \bibinfo{person}{L{\'a}szl{\'o} Szirmay-Kalos}, \bibinfo{person}{Gy{\"o}rgy
  Antal}, {and} \bibinfo{person}{Ferenc Csonka}.}
  \bibinfo{year}{2002}\natexlab{}.
\newblock \showarticletitle{A simple and robust mutation strategy for the
  {M}etropolis light transport algorithm}. In
  \bibinfo{booktitle}{\emph{Computer Graphics Forum}},
  Vol.~\bibinfo{volume}{21}. Wiley Online Library, \bibinfo{pages}{531--540}.
\newblock


\bibitem[Kilgariff et~al\mbox{.}(2018)]%
        {Kilgariff:2018:Nvidia}
\bibfield{author}{\bibinfo{person}{Emmett Kilgariff}, \bibinfo{person}{Henry
  Moreton}, \bibinfo{person}{Nick Stam}, {and} \bibinfo{person}{Brandon Bell}.}
  \bibinfo{year}{2018}\natexlab{}.
\newblock \showarticletitle{NVIDIA Turing Architecture In-Depth}.
\newblock \bibinfo{journal}{\emph{URL: https://devblogs. nvidia.
  com/nvidia-turing-architecture-indepth/(visited on 2020-05-11)}}
  (\bibinfo{year}{2018}).
\newblock


\bibitem[Kozlowski and Cheblokov(2021)]%
        {Pawel:2019:ReLAX}
\bibfield{author}{\bibinfo{person}{Pawel Kozlowski} {and} \bibinfo{person}{Tim
  Cheblokov}.} \bibinfo{year}{2021}\natexlab{}.
\newblock \showarticletitle{ReLAX: A Denoiser Tailored to Work with the ReSTIR
  Algorithm}.
\newblock \bibinfo{journal}{\emph{GPU Technology Conference}}
  (\bibinfo{year}{2021}).
\newblock


\bibitem[Lafortune and Willems(1993)]%
        {Lafortune:1993:Bidirectional}
\bibfield{author}{\bibinfo{person}{Eric~P. Lafortune} {and}
  \bibinfo{person}{Yves~D. Willems}.} \bibinfo{year}{1993}\natexlab{}.
\newblock \showarticletitle{Bi-Directional Path Tracing}. In
  \bibinfo{booktitle}{\emph{Proceedings of 3rd International Conference on
  Computational Graphics and Visualization Techniques}}.
  \bibinfo{pages}{145--153}.
\newblock


\bibitem[Lai et~al\mbox{.}(2015)]%
        {Lai:2015:PMCDI}
\bibfield{author}{\bibinfo{person}{Yu-Chi Lai}, \bibinfo{person}{Hsuan-Ting
  Chou}, \bibinfo{person}{Kuo-Wei Chen}, {and} \bibinfo{person}{Shaohua Fan}.}
  \bibinfo{year}{2015}\natexlab{}.
\newblock \showarticletitle{Robust and efficient adaptive direct lighting
  estimation}.
\newblock \bibinfo{journal}{\emph{The Visual Computer}} \bibinfo{volume}{31},
  \bibinfo{number}{1} (\bibinfo{year}{2015}), \bibinfo{pages}{83--91}.
\newblock


\bibitem[Lai and Dyer(2007)]%
        {Lai:2007:PMCPT}
\bibfield{author}{\bibinfo{person}{Yu-Chi Lai} {and} \bibinfo{person}{Charles
  Dyer}.} \bibinfo{year}{2007}\natexlab{}.
\newblock \bibinfo{booktitle}{\emph{Population Monte Carlo Path Tracing}}.
\newblock \bibinfo{type}{{T}echnical {R}eport}.
  \bibinfo{institution}{University of Wisconsin-Madison Department of Computer
  Sciences}.
\newblock


\bibitem[Lai et~al\mbox{.}(2007)]%
        {Lai:2007:PMCERPT}
\bibfield{author}{\bibinfo{person}{Yu-Chi Lai}, \bibinfo{person}{Shao~Hua Fan},
  \bibinfo{person}{Stephen Chenney}, {and} \bibinfo{person}{Charcle Dyer}.}
  \bibinfo{year}{2007}\natexlab{}.
\newblock \showarticletitle{Photorealistic image rendering with population
  {M}onte {C}arlo energy redistribution}. In
  \bibinfo{booktitle}{\emph{Proceedings of the 18th Eurographics conference on
  Rendering Techniques}}. \bibinfo{pages}{287--295}.
\newblock


\bibitem[Lai et~al\mbox{.}(2009)]%
        {Lai:2009:PMCTemporal}
\bibfield{author}{\bibinfo{person}{Yu-Chi Lai}, \bibinfo{person}{Feng Liu},
  {and} \bibinfo{person}{Charles Dyer}.} \bibinfo{year}{2009}\natexlab{}.
\newblock \bibinfo{booktitle}{\emph{Physically-based animation rendering with
  {M}arkov chain {M}onte {C}arlo}}.
\newblock \bibinfo{type}{{T}echnical {R}eport}.
  \bibinfo{institution}{University of Wisconsin-Madison Department of Computer
  Sciences}.
\newblock


\bibitem[Lehtinen et~al\mbox{.}(2013)]%
        {Lehtinen:2013:GDMLT}
\bibfield{author}{\bibinfo{person}{Jaakko Lehtinen}, \bibinfo{person}{Tero
  Karras}, \bibinfo{person}{Samuli Laine}, \bibinfo{person}{Miika Aittala},
  \bibinfo{person}{Fr{\'e}do Durand}, {and} \bibinfo{person}{Timo Aila}.}
  \bibinfo{year}{2013}\natexlab{}.
\newblock \showarticletitle{Gradient-domain {M}etropolis light transport}.
\newblock \bibinfo{journal}{\emph{ACM Transactions on Graphics (TOG)}}
  \bibinfo{volume}{32}, \bibinfo{number}{4} (\bibinfo{year}{2013}),
  \bibinfo{pages}{1--12}.
\newblock


\bibitem[Lin et~al\mbox{.}(2022)]%
        {Lin:2022:GRIS}
\bibfield{author}{\bibinfo{person}{Daqi Lin}, \bibinfo{person}{Markus
  Kettunen}, \bibinfo{person}{Benedikt Bitterli}, \bibinfo{person}{Jacopo
  Pantaleoni}, \bibinfo{person}{Cem Yuskel}, {and} \bibinfo{person}{Chris
  Wyman}.} \bibinfo{year}{2022}\natexlab{}.
\newblock \showarticletitle{Generalized Resampled Importance Sampling:
  Foundations of ReSTIR}.
\newblock \bibinfo{journal}{\emph{ACM Transactions on Graphics (TOG)}}
  \bibinfo{volume}{41}, \bibinfo{number}{75}.
\newblock


\bibitem[Lin et~al\mbox{.}(2021)]%
        {Lin:2021:ReSTIRVol}
\bibfield{author}{\bibinfo{person}{Daqi Lin}, \bibinfo{person}{Chris Wyman},
  {and} \bibinfo{person}{Cem Yuksel}.} \bibinfo{year}{2021}\natexlab{}.
\newblock \showarticletitle{Fast volume rendering with spatiotemporal reservoir
  resampling}.
\newblock \bibinfo{journal}{\emph{ACM Transactions on Graphics (TOG)}}
  \bibinfo{volume}{40}, \bibinfo{number}{6} (\bibinfo{year}{2021}),
  \bibinfo{pages}{1--18}.
\newblock


\bibitem[Metropolis et~al\mbox{.}(1953)]%
        {Metropolis:1953:MH}
\bibfield{author}{\bibinfo{person}{Nicholas Metropolis},
  \bibinfo{person}{Arianna~W Rosenbluth}, \bibinfo{person}{Marshall~N
  Rosenbluth}, \bibinfo{person}{Augusta~H Teller}, {and}
  \bibinfo{person}{Edward Teller}.} \bibinfo{year}{1953}\natexlab{}.
\newblock \showarticletitle{Equation of state calculations by fast computing
  machines}.
\newblock \bibinfo{journal}{\emph{The Journal of Chemical Physics}}
  \bibinfo{volume}{21}, \bibinfo{number}{6} (\bibinfo{year}{1953}),
  \bibinfo{pages}{1087--1092}.
\newblock


\bibitem[Mitchell(1987)]%
        {Mitchell:1987:antialiased}
\bibfield{author}{\bibinfo{person}{Don~P. Mitchell}.}
  \bibinfo{year}{1987}\natexlab{}.
\newblock \showarticletitle{Generating Antialiased Images at Low Sampling
  Densities}. In \bibinfo{booktitle}{\emph{Proceedings of the 14th Annual
  Conference on Computer Graphics and Interactive Techniques}}
  \emph{(\bibinfo{series}{SIGGRAPH '87})}. \bibinfo{publisher}{Association for
  Computing Machinery}, \bibinfo{address}{New York, NY, USA},
  \bibinfo{pages}{65–72}.
\newblock
\showISBNx{0897912276}
\urldef\tempurl%
\url{https://doi.org/10.1145/37401.37410}
\showDOI{\tempurl}


\bibitem[Neal(2001)]%
        {Neal:2001:AnnealedIS}
\bibfield{author}{\bibinfo{person}{Radford~M Neal}.}
  \bibinfo{year}{2001}\natexlab{}.
\newblock \showarticletitle{Annealed importance sampling}.
\newblock \bibinfo{journal}{\emph{Statistics and computing}}
  \bibinfo{volume}{11}, \bibinfo{number}{2} (\bibinfo{year}{2001}),
  \bibinfo{pages}{125--139}.
\newblock


\bibitem[NVIDIA(2017)]%
        {Nvidia:2017:Optix}
\bibfield{author}{\bibinfo{person}{NVIDIA}.} \bibinfo{year}{2017}\natexlab{}.
\newblock \bibinfo{title}{NVIDIA OptiX AI-Accelerated Denoiser}.
\newblock
  \bibinfo{howpublished}{\url{https://developer.nvidia.com/optix-denoiser}}.
\newblock


\bibitem[NVIDIA(2022)]%
        {Nvidia:2022:NRD}
\bibfield{author}{\bibinfo{person}{NVIDIA}.} \bibinfo{year}{2022}\natexlab{}.
\newblock \bibinfo{title}{NVIDIA Real-time denoisers (NRD)}.
\newblock
  \bibinfo{howpublished}{\url{https://developer.nvidia.com/rtx/ray-tracing/rt-denoisers}}.
\newblock


\bibitem[Otsu et~al\mbox{.}(2018)]%
        {Otsu:2018:Geometry}
\bibfield{author}{\bibinfo{person}{Hisanari Otsu}, \bibinfo{person}{Johannes
  Hanika}, \bibinfo{person}{Toshiya Hachisuka}, {and} \bibinfo{person}{Carsten
  Dachsbacher}.} \bibinfo{year}{2018}\natexlab{}.
\newblock \showarticletitle{Geometry-aware Metropolis light transport}.
\newblock \bibinfo{journal}{\emph{ACM Transactions on Graphics (TOG)}}
  \bibinfo{volume}{37}, \bibinfo{number}{6} (\bibinfo{year}{2018}),
  \bibinfo{pages}{1--11}.
\newblock


\bibitem[Ouyang et~al\mbox{.}(2021)]%
        {Ouyang:2021:ReSTIRGI}
\bibfield{author}{\bibinfo{person}{Yaobin Ouyang}, \bibinfo{person}{Shiqiu
  Liu}, \bibinfo{person}{Markus Kettunen}, \bibinfo{person}{Matt Pharr}, {and}
  \bibinfo{person}{Jacopo Pantaleoni}.} \bibinfo{year}{2021}\natexlab{}.
\newblock \showarticletitle{ReSTIR GI: Path Resampling for Real-Time Path
  Tracing}. In \bibinfo{booktitle}{\emph{Computer Graphics Forum}},
  Vol.~\bibinfo{volume}{40}. Wiley Online Library, \bibinfo{pages}{17--29}.
\newblock


\bibitem[Schied et~al\mbox{.}(2017)]%
        {Schied:2017:FilteringA}
\bibfield{author}{\bibinfo{person}{Christoph Schied}, \bibinfo{person}{Anton
  Kaplanyan}, \bibinfo{person}{Chris Wyman}, \bibinfo{person}{Anjul Patney},
  \bibinfo{person}{Chakravarty R~Alla Chaitanya}, \bibinfo{person}{John
  Burgess}, \bibinfo{person}{Shiqiu Liu}, \bibinfo{person}{Carsten
  Dachsbacher}, \bibinfo{person}{Aaron Lefohn}, {and} \bibinfo{person}{Marco
  Salvi}.} \bibinfo{year}{2017}\natexlab{}.
\newblock \showarticletitle{Spatiotemporal variance-guided filtering: real-time
  reconstruction for path-traced global illumination}.
\newblock In \bibinfo{booktitle}{\emph{Proceedings of High Performance
  Graphics}}. \bibinfo{pages}{1--12}.
\newblock


\bibitem[Schied et~al\mbox{.}(2018)]%
        {Schied:2018:FilteringB}
\bibfield{author}{\bibinfo{person}{Christoph Schied},
  \bibinfo{person}{Christoph Peters}, {and} \bibinfo{person}{Carsten
  Dachsbacher}.} \bibinfo{year}{2018}\natexlab{}.
\newblock \showarticletitle{Gradient estimation for real-time adaptive temporal
  filtering}.
\newblock \bibinfo{journal}{\emph{Proceedings of the ACM on Computer Graphics
  and Interactive Techniques}} \bibinfo{volume}{1}, \bibinfo{number}{2}
  (\bibinfo{year}{2018}), \bibinfo{pages}{1--16}.
\newblock


\bibitem[Talbot et~al\mbox{.}(2005)]%
        {Talbot:2005:Importance}
\bibfield{author}{\bibinfo{person}{Justin Talbot}, \bibinfo{person}{David
  Cline}, {and} \bibinfo{person}{Parris Egbert}.}
  \bibinfo{year}{2005}\natexlab{}.
\newblock \showarticletitle{Importance Resampling for Global Illumination}. In
  \bibinfo{booktitle}{\emph{Eurographics Symposium on Rendering (2005)}},
  \bibfield{editor}{\bibinfo{person}{Kavita Bala} {and} \bibinfo{person}{Philip
  Dutre}} (Eds.). \bibinfo{publisher}{The Eurographics Association}.
\newblock
\showISBNx{3-905673-23-1}
\showISSN{1727-3463}
\urldef\tempurl%
\url{https://doi.org/10.2312/EGWR/EGSR05/139-146}
\showDOI{\tempurl}


\bibitem[Talbot(2005)]%
        {Talbot:2005:RIS}
\bibfield{author}{\bibinfo{person}{Justin~F Talbot}.}
  \bibinfo{year}{2005}\natexlab{}.
\newblock \bibinfo{booktitle}{\emph{Importance resampling for global
  illumination}}.
\newblock \bibinfo{publisher}{Brigham Young University}.
\newblock


\bibitem[Van~de Woestijne et~al\mbox{.}(2017)]%
        {Van:2017:MLTTemporal}
\bibfield{author}{\bibinfo{person}{Joran Van~de Woestijne},
  \bibinfo{person}{Roald Frederickx}, \bibinfo{person}{Niels Billen}, {and}
  \bibinfo{person}{Philip Dutr{\'e}}.} \bibinfo{year}{2017}\natexlab{}.
\newblock \showarticletitle{Temporal coherence for {M}etropolis light
  transport}. In \bibinfo{booktitle}{\emph{Eurographics Symposium on
  Rendering-Experimental Ideas \& Implementations}}. Eurographics Association,
  \bibinfo{pages}{55--63}.
\newblock


\bibitem[Veach(1998)]%
        {Veach:1998:Robust}
\bibfield{author}{\bibinfo{person}{Eric Veach}.}
  \bibinfo{year}{1998}\natexlab{}.
\newblock \bibinfo{booktitle}{\emph{Robust Monte Carlo methods for light
  transport simulation}}.
\newblock \bibinfo{publisher}{Stanford University}.
\newblock


\bibitem[Veach and Guibas(1995a)]%
        {Veach:1995:Bidirectional}
\bibfield{author}{\bibinfo{person}{Eric Veach} {and} \bibinfo{person}{Leonidas
  Guibas}.} \bibinfo{year}{1995}\natexlab{a}.
\newblock \showarticletitle{Bidirectional estimators for light transport}.
\newblock In \bibinfo{booktitle}{\emph{Photorealistic Rendering Techniques}}.
  \bibinfo{publisher}{Springer}, \bibinfo{pages}{145--167}.
\newblock


\bibitem[Veach and Guibas(1995b)]%
        {Veach:1995:MIS}
\bibfield{author}{\bibinfo{person}{Eric Veach} {and}
  \bibinfo{person}{Leonidas~J Guibas}.} \bibinfo{year}{1995}\natexlab{b}.
\newblock \showarticletitle{Optimally combining sampling techniques for Monte
  Carlo rendering}. In \bibinfo{booktitle}{\emph{Proceedings of the 22nd Annual
  Conference on Computer Graphics and Interactive Techniques}}.
  \bibinfo{pages}{419--428}.
\newblock


\bibitem[Veach and Guibas(1997)]%
        {Veach:1997:MLT}
\bibfield{author}{\bibinfo{person}{Eric Veach} {and}
  \bibinfo{person}{Leonidas~J Guibas}.} \bibinfo{year}{1997}\natexlab{}.
\newblock \showarticletitle{Metropolis light transport}. In
  \bibinfo{booktitle}{\emph{Proceedings of the 24th annual conference on
  Computer graphics and interactive techniques}}. \bibinfo{pages}{65--76}.
\newblock


\bibitem[Ward et~al\mbox{.}(1988)]%
        {Ward:1988:Radiance}
\bibfield{author}{\bibinfo{person}{Gregory~J Ward}, \bibinfo{person}{Francis~M
  Rubinstein}, {and} \bibinfo{person}{Robert~D Clear}.}
  \bibinfo{year}{1988}\natexlab{}.
\newblock \showarticletitle{A ray tracing solution for diffuse
  interreflection}. In \bibinfo{booktitle}{\emph{Proceedings of the 15th Annual
  Conference on Computer Graphics and Interactive Techniques}}.
  \bibinfo{pages}{85--92}.
\newblock


\bibitem[Wyman(2021)]%
        {Wyman:2021:WRS}
\bibfield{author}{\bibinfo{person}{Chris Wyman}.}
  \bibinfo{year}{2021}\natexlab{}.
\newblock \showarticletitle{Weighted Reservoir Sampling: Randomly Sampling
  Streams}.
\newblock In \bibinfo{booktitle}{\emph{Ray Tracing Gems II}}.
  \bibinfo{publisher}{Springer}, \bibinfo{pages}{345--349}.
\newblock


\bibitem[Wyman and Panteleev(2021)]%
        {Wyman:2021:Rearchitecting}
\bibfield{author}{\bibinfo{person}{Chris Wyman} {and} \bibinfo{person}{Alexey
  Panteleev}.} \bibinfo{year}{2021}\natexlab{}.
\newblock \showarticletitle{Rearchitecting Spatiotemporal Resampling for
  Production}.
\newblock \bibinfo{journal}{\emph{ACM/Eurographics Symposium on High
  Performance Graphics}} (\bibinfo{year}{2021}).
\newblock


\end{thebibliography}



\begin{thebibliography}{1}


\ifx \showCODEN    \undefined \def \showCODEN     #1{\unskip}     \fi
\ifx \showDOI      \undefined \def \showDOI       #1{#1}\fi
\ifx \showISBNx    \undefined \def \showISBNx     #1{\unskip}     \fi
\ifx \showISBNxiii \undefined \def \showISBNxiii  #1{\unskip}     \fi
\ifx \showISSN     \undefined \def \showISSN      #1{\unskip}     \fi
\ifx \showLCCN     \undefined \def \showLCCN      #1{\unskip}     \fi
\ifx \shownote     \undefined \def \shownote      #1{#1}          \fi
\ifx \showarticletitle \undefined \def \showarticletitle #1{#1}   \fi
\ifx \showURL      \undefined \def \showURL       {\relax}        \fi
\providecommand\bibfield[2]{#2}
\providecommand\bibinfo[2]{#2}
\providecommand\natexlab[1]{#1}
\providecommand\showeprint[2][]{arXiv:#2}

\bibitem[Lin et~al\mbox{.}(2022)]%
        {Lin:2022:GRIS}
\bibfield{author}{\bibinfo{person}{Daqi Lin}, \bibinfo{person}{Markus
  Kettunen}, \bibinfo{person}{Benedikt Bitterli}, \bibinfo{person}{Jacopo
  Pantaleoni}, \bibinfo{person}{Cem Yuskel}, {and} \bibinfo{person}{Chris
  Wyman}.} \bibinfo{year}{2022}\natexlab{}.
\newblock \showarticletitle{Generalized Resampled Importance Sampling:
  Foundations of ReSTIR}.
\newblock \bibinfo{journal}{\emph{ACM Transactions on Graphics (TOG)}}
  \bibinfo{volume}{41}, \bibinfo{number}{75}.
\newblock


\end{thebibliography}

\appendix

\section{Unbiased Contribution Weights and Elimination of Startup Bias}
\label{app:StartupBiasElimination}

\eqref{ModifiedContributionWeight} shows how to update the contribution weight $W(x^k)$ of a mutated sample $x^k$ from the Markov chain $x^0, ..., x^k, ...$, generated with target function $\phat$. Here we prove this rule yields an unbiased contribution weight for any mutated sample $x^k$, \ie, for any $f$ with the same or smaller support,
\begin{equation}
    \label{eq:UnbiasedContributionWeight}
    \mathbb{E}[f(x^k) W(x^k)] = \int_{\Omega} f(x)\ \diff x.
\end{equation}
We assume sample $x^0$ initializing the chain has the same support $\Omega$ as target function $\phat$. For us, this is guaranteed by chained applications of RIS with a valid shift map in \algref{TemporalReuse}. Any $x^0$ chosen by RIS is not distributed exactly proportional to $\phat$ (unless we have infinite samples), however its contribution weight $W(x^0)$ is unbiased and satisfies \eqref{UnbiasedContributionWeight} \citep{Lin:2022:GRIS}. Next, we show access to $W(x^0)$ is sufficient to eliminate any startup bias with MH.

\paragraph{Proof} To show \eqref{UnbiasedContributionWeight} holds, we first express the update rule for contribution weight $W(x^k)$ (for $k > 0$) in terms of the previous sample $x^{k-1}$ in the chain, as follows:
\begin{equation}
    \label{eq:ModifiedContributionWeightPrevSample}
    W(x^k) = \frac{\phat(x^{k-1})}{\phat(x^k)} W(x^{k-1}).
\end{equation}
This is equivalent to \eqref{ModifiedContributionWeight}, shown by recursively unfolding this relationship for all prior samples $x^{k-1}$ to $x^1$ in the chain. As in \secref{MetropolisHastings}, we also assume a candidate mutation to $x^{k-1}$ is generated using the proposal density $T(x^{k-1} \rightarrow z^{k-1})$, with acceptance probability for candidate $z^{k-1}$ given by \eqref{MHAcceptance}:
\begin{equation}
    a(x^{k-1} \rightarrow z^{k-1}) := \min\left(1, \frac{\phat(z^{k-1})\ T(z^{k-1} \rightarrow x^{k-1})}{\phat(x^{k-1})\ T(x^{k-1} \rightarrow z^{k-1})}\right).
\end{equation}
Metropolis--Hastings sets $x^k = z^{k-1}$ with probability $a$; otherwise $x^k = x^{k-1}$. This lets us rewrite the expectation in \eqref{UnbiasedContributionWeight}:
\begin{align}
\mathbb{E}[f(x^k) W(x^k)] = &\mathbb{E}\left[f(x^k) \frac{\phat(x^{k-1})}{\phat(x^k)} W(x^{k-1})\right]\nonumber\\
= &\mathbb{E}\left[f(z^{k-1}) a(x^{k-1} \rightarrow z^{k-1}) \frac{\phat(x^{k-1})}{\phat(z^{k-1})} W(x^{k-1})\right]\nonumber\\ 
&+\ \mathbb{E}[f(x^{k-1}) (1 - a(x^{k-1} \rightarrow z^{k-1})) W(x^{k-1})].
\end{align}
Rearranging the terms slightly yields
\begin{align}
    \label{eq:ExpectationExpanded}
    \mathbb{E}[f(x^k) W(x^k)] =\ &\mathbb{E}[f(x^{k-1}) W(x^{k-1})]\ +\nonumber\\
&\mathbb{E}\left[f(z^{k-1}) a(x^{k-1} \rightarrow z^{k-1}) \frac{\phat(x^{k-1})}{\phat(z^{k-1})} W(x^{k-1})\right]\ -\nonumber\\ 
&\mathbb{E}[f(x^{k-1}) a(x^{k-1} \rightarrow z^{k-1}) W(x^{k-1})].
\end{align}
We now write each expectation as an integral. First, assume an inductive hypothesis $\mathbb{E}[f(x^{k-1}) W(x^{k-1})]\!=\! \int_{\Omega} f(x)\ \diff x$ for the $k\!-\!1^{\textrm{st}}$ MH iteration. Base case $k\!=\!1$ holds trivially, as $W(x^0)$ is an unbiased contribution weight. Next, note that for any integrable function $g(x^{k-1}, z^{k-1})$,
its expectation $\mathbb{E}[g(x^{k-1}, z^{k-1}) W(x^{k-1})]$ can be rewritten as a conditional expectation over candidate mutations:
\begin{multline}
    \mathbb{E}[g(x^{k-1}, z^{k-1}) W(x^{k-1})] = \mathbb{E}[E[g(x^{k-1}, z^{k-1}) \mid x^{k-1}]\  W(x^{k-1})]\\
    = \mathbb{E}\left[\left(\int_{\Omega} g(x^{k-1}, z) T(x^{k-1} \rightarrow z)\ \diff z\right) W(x^{k-1})\right],
\end{multline}
where $T$ is the proposal density used for mutations. This lets us expand out \eqref{ExpectationExpanded} as follows:
\begin{align}
    \label{eq:DoubleIntegral}
    \mathbb{E}[f(x^k) W(x^k)] =\ &\int_{\Omega} f(x)\ \diff x\ +\nonumber\\
    &\int_{\Omega} \int_{\Omega} f(z) a(x \rightarrow z) \frac{\phat(x)}{\phat(z)} T(x \rightarrow z)\ \diff z\ \diff x\ -\nonumber\\
    &\int_{\Omega} \int_{\Omega} f(x) a(x \rightarrow z) T(x \rightarrow z)\ \diff z\ \diff x.
\end{align}
Finally, we show that the two double integrals cancel each other out (resulting in $\mathbb{E}[f(x^k) W(x^k)] = \int_{\Omega} f(x)\ \diff x$) by invoking the detailed balance condition from \eqref{DetailedBalance}:
\begin{equation}
    \phat(x) T(x \rightarrow z) a(x \rightarrow z) = \phat(z) T(z \rightarrow x) a(z \rightarrow x)
\end{equation}
and rewriting it as:
\begin{equation}
    T(x \rightarrow z) a(x \rightarrow z) = \frac{\phat(z)}{\phat(x)} T(z \rightarrow x) a(z \rightarrow x).
\end{equation}
Substituting for $T(x \rightarrow z) a(x \rightarrow z)$ in the third line of \eqref{DoubleIntegral} yields the same integral as the second line, but with integration variables $x$ and $z$ swapped. Renaming $x$ and $z$ and swapping the integration order in the third line allows cancellation, simplifying to $\int_{\Omega} f(x)\ \diff x$, yielding \eqref{UnbiasedContributionWeight} and giving a proof by induction. 

\section{MIS Weights for Temporal Reuse}
\label{app:MISTemporal}

ReSTIR uses MIS weights during resampling (Equations \ref{eq:ResamplingWeightMIS} and \ref{eq:ResamplingWeightShift}) to mitigate noise and bias from reusing samples across pixels. We provide explicit expressions for the MIS weights used in \algref{TemporalReuse} here; \citet[Equations 37--38]{Lin:2022:GRIS} provide similar expressions for Pairwise MIS weights needed for spatial resampling.

Let $S_j\!:\!\Omega_j\!\rightarrow\!\Omega_i$ denote the shift map from pixel $j$ to pixel $i$. Let $x_i$ and $x_j$ further represent the corresponding samples for these pixels, and $S_j(x_j)\!=\!x_j'$ and $S_j^{-1}(x_i)\!=\!x_i'$ the respective shift mapped values. The MIS weights for $x_i$ and $x_j'$ are then given by:
\begin{align}
    m_i(x_i) = \frac{M_i\ \phat_i(x_i)}{M_i\ \phat_i(x_i) + M_j\ \phat_j(x_i')\ |\partial x_i' / \partial x_i|},\label{eq:MISTemporal1}\\
    m_j(x_j') = \frac{M_j\ \phat_j(x_j)\ |\partial x_j/\partial x_j'|}{M_j\ \phat_j(x_j)\ |\partial x_j/\partial x_j'| + M_i\ \phat_i(x_j')}\label{eq:MISTemporal2}.
\end{align}
We set $m_i(x_i) = 1$ and $m_j(x_j') = 0$ when valid shifts do not exist for $x_i$ and $x_j$ respectively; \citet[Section 5.6]{Lin:2022:GRIS} discusses properties of these MIS weights in detail.

\section{Transition Kernel for Mutating a Reconnection Vertex}
\label{app:TransitionKernelReconnection}

A mutation involving a reconnection vertex $\y_i$ requires modifying random numbers not just for $\y_i$, but also for  non-mutated vertices $\y_{i+1}$ and $\y_{i+2}$. This is because the solid angle PDFs used to sample outgoing directions $\omega'_{i}$ and $\omega_{i+1}$ depend on the mutated incoming directions $\omega'_{i-1}$ and $\omega'_{i}$, respectively. Here we derive \eqref{TransitionKernelRatio} by first noting the joint PDF for connecting mutated reconnection vertex $\y'_i$ to $\y'_{i+1}$ and $\y'_{i+1}$ to $\y'_{i+2}$ in the surface area measure is:
\begin{equation*}
    p(\y'_{i+2}, \y'_{i+1} | [\y'_0, \y'_1, \ldots, \y'_i]) = \delta(\y'_{i+2} - \y_{i+2}) \delta(\y'_{i+1} - \y_{i+1}).
\end{equation*}
This is a product of delta functions as $\y'_{i+1}=\y_{i+1}$ and $\y'_{i+2}=\y_{i+2}$ are the only valid vertex positions. 
In the PSS to path space mapping, the joint PDF for the mutated random numbers $\ubar'_{i+1}$ and $\ubar'_{i+2}$ (for vertices $\y_{i+1}$ and $\y_{i+2}$) is related to $ p(\y'_{i+2}, \y'_{i+1}\ |\ [\y'_0, \y'_1, \ldots, \y'_i])$ via a Jacobian determinant:
\begin{equation*}
    p(\ubar'_{i+2}, \ubar'_{i+1} | [\ubar'_0, \ubar'_1, \ldots, \ubar'_i]) = 
    p(\y'_{i+2}, \y'_{i+1} | [\y'_0, \y'_1, \ldots, \y'_i]) \jdet{\ybar'}{\ubar'}
\end{equation*}
This PDF serves as our proposal density $T(\ubar \rightarrow \ubar')$ for mutations, which then yields:
\begin{align}
    \frac{T(\ubar' \rightarrow \ubar)}{T(\ubar \rightarrow \ubar')} &= \jdet{\ubar'}{\ybar'} \jdet{\ybar}{\ubar}\frac{\delta(\y_{i+2} - \y'_{i+2}) \delta(\y_{i+1} - \y'_{i+1})}{\delta(\y'_{i+2} - \y_{i+2}) \delta(\y'_{i+1} - \y_{i+1})} \nonumber\\
    &= \jdet{\ubar'}{\bar{\omega}'} \jdet{\bar{\omega}'}{\ybar'}\jdet{\ybar}{\bar{\omega}}\jdet{\bar{\omega}}{\ubar}\nonumber\\
    &= \jdet{\bar{\omega}'}{\ybar'}\jdet{\ybar}{\bar{\omega}}\frac{p(\omega_{i-1}', \omega_i')}{p(\omega_{i-1}, \omega_i)} \frac{p(\omega_i', \omega_{i+1})}{p(\omega_i, \omega_{i+1})}\nonumber\\
    &= \frac{|\text{cos}\ \theta'|}{|\text{cos}\ \theta|} \frac{|\y_{i+1} - \y_{i}|^2}{|\y_{i+1} - \y'_{i}|^2} \frac{p(\omega_{i-1}', \omega_i')}{p(\omega_{i-1}, \omega_i)} \frac{p(\omega_i', \omega_{i+1})}{p(\omega_i, \omega_{i+1})}.
\end{align}
The delta functions in the first line cancel since they are symmetric. In the third line, we use the fact that the Jacobian determinant of a sampling scheme is the same as its inverse PDF \citep[Section 2]{Kelemen:2002:PSSMLT}. The final step substitutes in the definition of the Jacobians relating the solid angle and area measures.

\end{document}


\title{Supplemental Document for Decorrelating ReSTIR Samplers via MCMC Mutations} 

\author{Rohan Sawhney}
\email{rohansawhney@cs.cmu.edu}
\affiliation{%
  \institution{Carnegie Mellon University}
  \country{USA}
}
 
\author{Daqi Lin}
\email{daqil@nvidia.com}
\affiliation{%
  \institution{NVIDIA}
  \country{USA}
}

\author{Markus Kettunen}
\email{mkettunen@nvidia.com}
\affiliation{%
  \institution{NVIDIA}
  \country{Finland}
}

\author{Benedikt Bitterli}
\email{bbitterli@nvidia.com}
\affiliation{%
  \institution{NVIDIA}
  \country{USA}
}

\author{Ravi Ramamoorthi}
\email{ravir@cs.ucsd.edu}
\affiliation{%
  \institution{UC San Diego and NVIDIA}
  \country{USA}
}

\author{Chris Wyman}
\email{cwyman@nvidia.com}
\affiliation{%
  \institution{NVIDIA}
  \country{USA}
}

\author{Matt Pharr}
\email{mpharr@nvidia.com}
\affiliation{%
  \institution{NVIDIA}
  \country{USA}
}

%
%

\ccsdesc[500]{Computing methodologies~Ray Tracing}

%
%

\keywords{real-time rendering, resampled importance sampling, weighted reservoir sampling, Markov chain Monte Carlo}



\maketitle

\section{Why Mutations Decrease Covariance}

\newcommand{\Xto}[2]{X_{#2}^{#1}}
\newcommand{\Yto}[2]{Y_{#2}^{#1}}
\newcommand{\Zto}[1]{Z^{#1}}
\newcommand{\Wto}[2]{W_{#2}^{#1}}
\newcommand{\WYto}[2]{{W'}_{#2}^{#1}}
\newcommand{\WZto}[1]{W^{#1}}
\newcommand{\wto}[2]{w_{#2}^{#1}}
\newcommand{\pbarTo}[2]{\pbar_{#2}^{#1}}
\newcommand{\phatTo}[2]{\phat_{#2}^{#1}}
\newcommand{\phatAt}[1]{\phat^{#1}}
\newcommand{\pbarAt}[1]{\pbar^{#1}}

\newcommand{\TWZto}[1]{\tilde{W}^{#1}}
\newcommand{\Twto}[2]{\tilde{w}_{#2}^{#1}}
\newcommand{\TZto}[1]{\tilde{Z}^{#1}}

\newcommand{\fAt}[1]{f^{#1}}

\begin{table}[t]
    \begin{footnotesize}
    \centering
    \caption{Summary of notation}
    \begin{tabular}{c|l}
    \hline
        $\Xto{i}{j}$, $\Wto{i}{j}$    & Pixel $i$'s original input sample $j$ and its contribution weight \\
        $\Yto{i}{j}$, $\WYto{i}{j}$   & Pixel $i$'s mutated input sample $j$ and its contribution weight \\
        $\Zto{i}, \WZto{i}$           & Sample chosen for pixel $i$ from the $\Yto{i}{j}$ and its contribution weight \\
        $\wto{i}{j}$                  & Resampling weight for choosing $\Yto{i}{j}$ as the new $\Zto{i}$ \\
        $\phatAt{i}$, $\pbarAt{i}$    & Target function of pixel $i$ and its normalized target PDF \\
        $\phatTo{i}{j}$, $\pbarTo{i}{j}$ & Target function of pixel $i$'s input $j$ and its target PDF \\
        $\fAt{i}$                     & The integrand in pixel $i$, here $\fAt{i} = \phatAt{i}$ \\
        $\TZto{i}, \TWZto{i}$         & Sample chosen for pixel $i$ from the $\Xto{i}{j}$ and its contribution weight \\
        $\Twto{i}{j}$                 & Resampling weight for choosing $\Xto{i}{j}$ as the new $\TZto{i}$ \\
        $\|\cdot\|$                   & The 1-norm, $\|g\| = \int_{\Omega} \left|g(x)\right| \d{x}$, e.g., $\pbar = \phat / \|\phat\|$ \\
    \hline
    \end{tabular}
    \end{footnotesize}
    \label{tab:symbolSummary}
\end{table}

We present an argument for why mutating samples after temporal reuse reduces covariance in the final image. We study the problem in a simplified setting, which we believe captures the essence. 

In particular, we study the covariance between two pixels with target functions $\phatAt{1}$ and $\phatAt{2}$, with pixels 1 and 2 spatially reusing samples from random neighbors. We model random neighbor selection as follows: the input samples for pixel $i \in \{1, 2\}$ are $\Xto{i}{j}$ where $j$ ranges from 1 to $M$, and their contribution weights are $\Wto{i}{j}$. Superscripts denote the target pixel and subscripts the index of its $j$'th input; input samples for the pixels are assumed to be distinct. As in ReSTIR DI, we assume samples lie in the same domain $\Omega$ and share the support (i.e., agree on visibility), with shift mappings not needed for reuse between pixels (i.e., the identity shift mapping is used with light vertices as they are). We also use constant MIS weights $1/M$. With this setup, we show a reduction in covariance in the limit case as the number of mutations approaches infinity (assuming good importance sampling), and argue that benefits in the finite case arise from approximating the limit case.

We interpret the input samples $\Xto{i}{j}$ as the results of temporal resampling, and denote the final mutation results before spatial reuse $\Yto{i}{j}$ and their contribution weights $\WYto{i}{j}$. The assumption of good importance sampling means that $\Wto{i}{j} \approx 1/\pbarTo{i}{j}(\Xto{i}{j})$, where $\pbarTo{i}{j}$ is the normalized version of the target function $\phatTo{i}{j}$, i.e., $\pbarTo{i}{j} = \phatTo{i}{j} / \|\phatTo{i}{j}\|$. We again denote the target pixel in the superscript and the index of its $j$'th input pixel in the subscript. Assuming a large number of mutations implies that we can treat the mutation results $\Yto{i}{j}$ as independent of each other, the input samples $\Xto{i}{j}$, and their contribution weights $\Wto{i}{j}$.
The contribution weight of a mutated sample $\Yto{i}{j}$ is
\begin{equation}
    \WYto{i}{j} = \frac{\phatTo{i}{j}(\Xto{i}{j})}{\phatTo{i}{j}(\Yto{i}{j})} \Wto{i}{j},
\end{equation}
while the resampling weight for choosing sample $\Yto{i}{j}$ for pixel $i$ is
\begin{align}
    \wto{i}{j} & = \frac{1}{M} \phatAt{i}(\Yto{i}{j}) \WYto{i}{j} \\
     & = \frac{1}{M} \frac{\phatAt{i}(\Yto{i}{j})}{\phatTo{i}{j}(\Yto{i}{j})} \phatTo{i}{j}(\Xto{i}{j}) \Wto{i}{j}.
\end{align}
The sample $\Zto{i}$ selected for pixel $i$ by resampling proportionally to $\wto{i}{j}$ has the contribution weight
\begin{align}
    \WZto{i} & = \frac{1}{\phatAt{i}(\Zto{i})} \sum_{j=1}^M \wto{i}{j} \\
    & = \frac{1}{\phatAt{i}(\Zto{i})} \sum_{j=1}^M \frac{1}{M} 
    \frac{\phatAt{i}(\Yto{i}{j})}{\phatTo{i}{j}(\Yto{i}{j})} \phatTo{i}{j}(\Xto{i}{j}) \Wto{i}{j}.
\end{align}
Using the integrand of the rendering equation $\fAt{i}$ as our target function $\phatAt{i}$, the pixel estimate $\fAt{i}(\Zto{i}) \WZto{i}$ is
\begin{equation}
    \phatAt{i}(\Zto{i}) \WZto{i} = \frac{1}{M} \sum_{j=1}^M \frac{\phatAt{i}(\Yto{i}{j})}{\phatTo{i}{j}(\Yto{i}{j})} \phatTo{i}{j}(\Xto{i}{j}) \Wto{i}{j}, 
\end{equation}
and the covariance between estimates for pixels 1 and 2 becomes
\begin{multline}
    \label{eq:supp-covariance-of-two}
    \Cov\left( \phatAt{1}(\Zto{1}) \WZto{1}, \phatAt{2}(\Zto{1}) \WZto{2} \right) \\
    = \frac{1}{M^2} \sum_{j=1}^M \sum_{k=1}^M \Cov\left(  \frac{\phatAt{1}(\Yto{1}{j})}{\phatTo{1}{j}(\Yto{1}{j})} {\phatTo{1}{j}(\Xto{1}{j})} \Wto{1}{j},  \frac{\phatAt{2}(\Yto{2}{k})}{\phatTo{2}{k}(\Yto{2}{k})} {\phatTo{2}{k}(\Xto{2}{k})} \Wto{2}{k}
    \right) .
\end{multline}
Since we study the limit case in which the mutated samples $\Yto{i}{j}$ are independent of other random variables, we can rewrite this expression using the relation $\Cov(XY, Z) = \E[X] \Cov(Y, Z)$, which assumes that $X$ is independent of $Y$ and $Z$. Applying this relation to both parameters of the covariance yields
\begin{equation}
    \label{eq:supp-covariance-of-two-b}
    = \frac{1}{M^2} \sum_{j=1}^M \sum_{k=1}^M
        \E\left[ \frac{\phatAt{1}(\Yto{1}{j})}{\phatTo{1}{j}(\Yto{1}{j})} \right]
        \E\left[ \frac{\phatAt{2}(\Yto{2}{k})}{\phatTo{2}{k}(\Yto{2}{k})} \right]
        \Cov\left( \phatTo{1}{j}(\Xto{1}{j}) \Wto{1}{j}, \phatTo{2}{k}(\Xto{2}{k}) \Wto{2}{k}
    \right) .
\end{equation}
We now simplify the expression above by first writing $\E\left[\frac{\phatAt{i}(\Yto{i}{j})}{\phatTo{i}{j}(\Yto{i}{j})}\right]$ as $\frac{\|\phatAt{i}\|}{\|\phatTo{i}{j}\|} \E\left[\frac{\pbarAt{i}(\Yto{i}{j})}{\pbarTo{i}{j}(\Yto{i}{j})}\right]$, where we move the norms of the target functions outside the expectation. Since we study the limit case of near-infinite mutations, $\Yto{i}{j}$ has PDF $\pbarTo{i}{j}(\Yto{i}{j})$, which is the denominator in the expectation.
As a result, the expectation evaluates to the integral of the PDF in the numerator (i.e., 1), simplifying the expression for covariance to
\begin{equation}
    = \frac{\|\phatAt{1}\| \|\phatAt{2}\|}{M^2} \sum_{j=1}^M \sum_{k=1}^M  \Cov\left( \pbarTo{1}{j}(\Xto{1}{j}) \Wto{1}{j}, \pbarTo{2}{k}(\Xto{2}{k}) \Wto{2}{k} \right).
\end{equation}
Here the norms of $\phatTo{1}{j}$ and $\phatTo{2}{k}$ are used to normalize $\phat$ into $\pbar$ inside the covariance. Finally, we use the definition of covariance, $\Cov(X, Y) = \E[(X - \mu_X)(Y - \mu_Y)]$, and the relation $\E[\pbarTo{i}{j}(\Xto{i}{j}) \Wto{i}{j}] = \int_{\Omega} \pbarTo{i}{j}(x) \d{x} = 1$ (by the way unbiased contribution weights transform expectations into integrals) to evaluate $\mu_X$ and $\mu_Y$ and reach the final form for the covariance of the pixel estimates,
\begin{multline}
    \label{eq:supp-cov1}
    = \frac{\|\phatAt{1}\| \|\phatAt{2}\|}{M^2} \sum_{j=1}^M \sum_{k=1}^M \E\left[ (\pbarTo{1}{j}(\Xto{1}{j}) \Wto{1}{j} - 1)(\pbarTo{2}{k}(\Xto{2}{k}) \Wto{2}{k} - 1) \right] .
\end{multline}
This final expression for the covariance shows that when the input samples $\Xto{i}{j}$ are importance sampled well at their original pixels, i.e., $\Wto{i}{j} \approx 1 / \pbarTo{i}{j}(\Xto{i}{j})$, then both factors in the expectation tend to be small, yielding a small covariance as well.

We now consider the case without mutations, deriving covariance when $\TZto{1}$ and $\TZto{2}$ are resampled directly from the samples $\Xto{i}{j}$ without mutations. The resampling weights are
\begin{equation}
    \Twto{i}{j} = \frac{1}{M} \phatAt{i}(\Xto{i}{j}) \Wto{i}{j},
\end{equation}
while the chosen sample $\TZto{i}$ has contribution weight
\begin{equation}
    \TWZto{i} = \frac{1}{\phatAt{i}(\TZto{i})} \sum_{j=1}^M \Twto{i}{j} = 
    \frac{1}{\phatAt{i}(\TZto{i})} \sum_{j=1}^M \frac{1}{M} \phatAt{i}(\Xto{i}{j}) \Wto{i}{j}.
\end{equation}
We again set $\phatAt{i} = \fAt{i}$, which yields the pixel contribution
\begin{equation}
    \phatAt{i}(\TZto{i}) \TWZto{i} = \frac{1}{M} \sum_{j=1}^M \phatAt{i}(\Xto{i}{j}) \Wto{i}{j}.
\end{equation}
The pixel covariance then is
\begin{multline}
    \Cov\left( \phatAt{1}(\TZto{1}) \TWZto{1}, \phatAt{2}(\TZto{2}) \TWZto{2} \right) \\
    = \frac{1}{M^2} \sum_{j=1}^M \sum_{k=1}^M \Cov\left( \phatAt{1}(\Xto{1}{j}) \Wto{1}{j},  \phatAt{2}(\Xto{2}{k}) \Wto{2}{k} \right) ,
\end{multline}
which we simplify to
\begin{multline}
    = \frac{\|\phatAt{1}\| \|\phatAt{2}\|}{M^2} \sum_{j=1}^M \sum_{k=1}^M \Cov\left( \pbarAt{1}(\Xto{1}{j}) \Wto{1}{j},  \pbarAt{2}(\Xto{2}{k}) \Wto{2}{k} \right).
\end{multline}
As before, we finally express covariance with expectations,
\begin{multline}
    \label{eq:supp-cov2}
    = \frac{\|\phatAt{1}\| \|\phatAt{2}\|}{M^2} \sum_{j=1}^M \sum_{k=1}^M 
    \E\left[ (\pbarAt{1}(\Xto{1}{j}) \Wto{1}{j} - 1)(\pbarAt{2}(\Xto{2}{k}) \Wto{2}{k} - 1) \right] .
\end{multline}
We immediately observe the critical difference between \eqref{supp-cov1} and \eqref{supp-cov2}: the inputs $\Xto{i}{j}$ are in both cases distributed approximately proportionally to $\pbarTo{i}{j}$, where with good importance sampling we have $\Wto{i}{j} \approx 1 / \pbarTo{i}{j}(\Xto{i}{j})$; without mutations, the expressions inside the expectations approximately equal
\begin{equation}
    \left( \frac{\pbarAt{1}(\Xto{1}{j})}{\pbarTo{1}{j}(\Xto{1}{j})} - 1 \right)
    \left( \frac{\pbarAt{2}(\Xto{2}{k})}{\pbarTo{2}{j}(\Xto{2}{k})} - 1 \right),
\end{equation}
which have no guarantees of being small if the target functions of the pixels and their inputs are not similar. In fact, these ratios may be arbitrarily large. However, by applying many mutations, these expressions typically become much smaller, yielding
\begin{equation}
    \left( \frac{\pbarTo{1}{j}(\Xto{1}{j})}{\pbarTo{1}{j}(\Xto{1}{j})} - 1 \right)
    \left( \frac{\pbarTo{2}{k}(\Xto{2}{k})}{\pbarTo{2}{k}(\Xto{2}{k})} - 1 \right) \approx 0 ,
\end{equation}
This implies that mutations help bring covariance closer to zero even when the target functions are different.

We have demonstrated that with well-distributed input samples, a large number of mutations help decorrelate the inputs to reservoir sampling during spatial reuse, effectively making input paths independent. The resulting independence of the mutated paths 
suggests that covariance results merely from bad importance sampling of the original pixels, not from incompatibility between close-by pixels. 

Pixels whose covariance is minimized should share as few input samples as possible---the proof requires a complete separation. In practice, we only apply a small number of mutations, which does not completely decorrelate the input samples. We also randomize the input pixels for each target pixel, leading to a small probability of overlap between input samples. While we cannot realize the ideal in practice, we aim for it as much as possible---every mutation is a step closer to the limit, and heuristically we expect some decrease in covariance with a smaller number of mutations.

Covariance does not go to zero simply by increasing the number of mutations. We are still bounded by random overlaps between input pixels and the quality of importance sampling as seen in \eqref{supp-cov1}.

In practice, proper MIS weights also protect the renderer from the most terrible correlations and can be used to guarantee eventual convergence (see \citet{Lin:2022:GRIS}). A performance-optimized implementation that greedily neglects MIS weights is however much more prone to high covariance---this analysis predicts that mutations are especially effective at removing correlation artifacts in such use cases. It also predicts that covariance will be present especially between close-by pixels with very different target functions. ReSTIR implementations (such as ours) try to defend against such cases by using expensive MIS weights and/or careful neighbor selection; this analysis, together with our empirical results, suggests that mutations are a useful addition for further robustness.

\section{Why Mutations Do Not Decrease Variance (Much)}

Variance can be studied as a pixel's covariance with itself, $\Var(X) = \Cov(X, X)$. We do not prove that variance reduction cannot happen when mutations are used---in some cases, it can. However, as we noted earlier, the two pixels' input samples should be different to minimize covariance. This is not true in the case of variance: a pixel, tautologically, has the same input samples as itself, and the mechanism to reduce covariance does not apply to variance---the corresponding inputs have 100\% correlation. This is in line with our empirical findings: mutations have little impact on variance.

\bibliographystyle{ACM-Reference-Format}
\bibliography{RestirMCMC}